\documentclass[trackchanges,twocolumn]{aastex7}

\usepackage{amsmath,amssymb,graphicx,xcolor,makecell}

\newcommand{\fesclya}{$f_\text{esc}^{\text{Ly}\alpha}$}
\newcommand{\fesclyC}{$f_\text{esc}^{\text{LyC}}$}

\newcommand*\diff{\mathop{}\!\mathrm{d}}

\newcommand{\oiii}{[O\,{\sc iii}]}

\newcommand{\civ}{C\,{\sc iv}}

\newcommand{\hb}{H$\beta$}
\newcommand{\lya}{Ly$\alpha$}

\newcommand{\ewlya}{$\text{EW}^\text{rest}_{\text{Ly}\alpha}$}

\begin{document}

\title{Escape fractions from unattenuated {\lya} emitters around luminous $z>6$ quasars}
\correspondingauthor{Minghao Yue}

\author[0000-0002-5367-8021]{Minghao Yue}
\email{myue@mit.edu}
\affiliation{MIT Kavli Institute for Astrophysics and Space Research, 77 Massachusetts Ave., Cambridge, MA 02139, USA}

\author[0000-0003-2895-6218]{Anna-Christina Eilers}
\email{eilers@mit.edu}
\affiliation{MIT Kavli Institute for Astrophysics and Space Research, 77 Massachusetts Ave., Cambridge, MA 02139, USA}

\author[0000-0003-2871-127X]{Jorryt Matthee}
\email{jorryt.matthee@ist.ac.at}
\affiliation{Institute of Science and Technology Austria (ISTA), Am Campus 1, 3400 Klosterneuburg, Austria}

\author[0000-0003-3997-5705]{Rohan P. Naidu}
\email{rnaidu@mit.edu}
\altaffiliation{NHFP Hubble Fellow}
\affiliation{MIT Kavli Institute for Astrophysics and Space Research, 77 Massachusetts Ave., Cambridge, MA 02139, USA}

\author[0000-0002-3120-7173]{Rongmon Bordoloi}
\email{rbordol@ncsu.edu}
\affiliation{Department of Physics, North Carolina State University, Raleigh, 27695, North Carolina, USA}

\author[0000-0003-0821-3644]{Frederick B. Davies}
\email{davies@mpia-hd.mpg.de}
\affiliation{Max Planck Institut f\"{u}r Astronomie, K\"{o}nigstuhl 17, D-69117 Heidelberg, Germany}

\author[0000-0002-7054-4332]{Joseph F. Hennawi}
\email{joe@physics.ucsb.edu}
\affiliation{Department of Physics, University of California, Santa Barbara, CA 93106-9530, USA}
\affiliation{Leiden Observatory, Leiden University, P.O. Box 9513, 2300 RA Leiden, The Netherlands}

\author[0000-0001-9044-1747]{Daichi Kashino}
\email{kashinod.astro@gmail.com}
\affiliation{National Astronomical Observatory of Japan, 2-21-1 Osawa, Mitaka, Tokyo 181-8588, Japan}

\author[0000-0003-0417-385X]{Ruari Mackenzie}
\email{mruari@phys.ethz.ch}
\affiliation{Institute of Physics, Laboratory of Astrophysics, Ecole Polytechnique F\'ed\'erale de Lausanne (EPFL),\and\small Observatoire de Sauverny,1290 Versoix, Switzerland.}

\author[0000-0003-3769-9559]{Robert A. Simcoe}
\email{simcoe@space.mit.edu}
\affiliation{MIT Kavli Institute for Astrophysics and Space Research, 77 Massachusetts Ave., Cambridge, MA 02139, USA}

\begin{abstract}

Ionized proximity zones around luminous quasars provide a unique laboratory to characterize the {\lya} emission lines from $z>6$ galaxies without significant attenuation from the intergalactic medium (IGM).
However, {\lya} line measurements for galaxies within high-redshift quasars' proximity zones have been rare so far.
Here we present deep spectroscopic observations obtained with the NIRSpec/MSA instrument on the James Webb Space Telescope (JWST) of galaxies in two $z>6$ quasar fields. 
We measure the {\lya} line fluxes for 50 galaxies at $6<z<7$ with UV absolute magnitude $M_\text{UV}<-19$ (median $M_\text{UV}=-19.97$), among which 15 are located near the luminous quasars, i.e. within $\Delta v<2500\rm\,km\,s^{-1}$. 
We find that galaxies near the quasars show significant flux bluewards of the systemic {\lya} wavelength, and have higher {\lya} equivalent width compared to galaxies at similar redshifts that are not located within the quasars' environment. 
Our result indicates little or no redshift evolution for the {\lya}-emitter fraction from $z\sim6.4$ to $z\sim5$.
Leveraging the low IGM opacity in the quasars' vicinity,
we evaluate the {\lya} escape fraction ({\fesclya}) of high-redshift galaxies. Our analysis suggests that galaxies at $\langle z\rangle\approx6.4$ have an average $f_\text{esc}^{\text{Ly}\alpha}=0.14\pm0.04$.  
This value is consistent with reionization models where the Lyman continuum escape fraction is low $(f_\text{esc}^\text{LyC}\lesssim0.1)$ for luminous galaxies, and where the most luminous galaxies have only a minor contribution to the total ionizing photon budget.

\end{abstract}


\section{Introduction} \label{sec:intro}

The Epoch of Reionization (EoR) represents the last major phase transition of the universe, where the neutral hydrogen fraction of the intergalactic medium (IGM) changes from $\sim1$ to $\sim0$ between redshift $z\sim20$ and $z\sim5.5$ \citep[e.g.,][]{bosman22,robertson22}. 
The overall history of reionization has been constrained by various observations, including the Cosmic Microwave Background measurements \citep[e.g.,][]{cmb}, quasar transmission fluxes and damping wings \citep[e.g.,][]{fan06,yang20b,jin23,durovcikova24}, and {\lya} lines from high-redshift galaxies \citep[e.g.,][]{mason18,mason19,wold22, tang24b, kageura25, umeda25}.
Nonetheless, it is still under debate which type of sources dominates the ionizing photon production.
Specifically, several studies have suggested that luminous galaxies make a major contribution to reionization \citep[e.g.,][]{sharma16,naidu20,matthee22}, while others found that the ionizing photon production is dominated by faint galaxies \citep[e.g.,][]{bouwens12,finkelstein19,atek24}. 

One reason for this debate is the highly uncertain Lyman continuum (LyC) escape fraction ({\fesclyC}) for the EoR galaxies. 
Due to the high IGM opacity to LyC photons at $z\gtrsim6$, it is impossible to directly measure the leaking LyC flux from EoR galaxies. 
Several indirect methods have been employed to infer the escape fraction of galaxies at $z\gtrsim6$, including observations for low-redshift counterparts of $z>6$ galaxies \citep[e.g.,][]{steidel18, naidu18, izotov18}, computing the ionizing photon budget that is needed to produce the observed neutral fraction evolution \citep[e.g.,][]{finkelstein19,naidu20,robertson22}, and adopting predictions from cosmological simulations \citep[e.g.,][]{ma15,ma16,Rosdahl22}. Nevertheless, the inferred LyC escape fraction shows large discrepancies,  ranging from $f_\text{esc}^\text{LyC}\lesssim5\%$ \citep[e.g.,][]{Chisholm22, mascia23, simmons24, papovich25} to $f_\text{esc}^\text{LyC}\gtrsim20\%$ \citep[e.g.,][]{naidu20,jaskot2024}.

At $4\lesssim z\lesssim5$, where the IGM is mostly opaque to LyC photons but has high transmission to {\lya} lines from galaxies \citep[e.g.,][]{meyer25}, the LyC escape fraction of galaxies can be estimated using the 
 escape fraction of the {\lya} emission line \citep[{\fesclya}; e.g.,][]{izotov22,begley24}. However, this method is not applicable to most galaxies in the EoR, as the {\lya} line is heavily attenuated by the neutral hydrogen in the $z>6$ IGM. The only viable approach to measure the unattenuated {\lya} lines from EoR galaxies is to target galaxies that reside in ionized bubbles, where the IGM is transparent to {\lya} photons \citep[e.g.,][]{matthee18, lu24,whitler24,Torralba24,chen25}. Such ionized bubbles are often produced by galaxy overdensities, and finding these ionized bubbles requires deep spectroscopy over wide fields \citep[e.g.,][]{nikolic25}.
 
In this context, proximity zones of high-redshift quasars offer a unique opportunity to investigate the unattenuated {\lya} properties of EoR galaxies \citep[e.g.,][]{bosman20,protusova24}. 
Quasars at $z>6$ are expected to produce large ionized bubbles around them, as implied by radiative transfer simulations \citep[e.g.,][]{davies2020b}.
Deep spectroscopy for galaxies residing in the proximity of high-redshift quasars will put direct constraints on the {\lya} line properties of these galaxies.

In this work, we present {\em James Webb Space Telescope (JWST)} NIRSpec/MSA spectroscopy for galaxies near two $z>6$ quasars, i.e. J0100+2802 $(z=6.327)$ and J1148+5251 $(z=6.42)$. We analyze the {\lya} line properties of these galaxies, and try to answer the following questions:
(1) For galaxies around high-redshift quasars, can we detect their {\lya} emission lines without significant IGM attenuation? (2) How can we use galaxies near $z>6$ quasars to constrain the {\lya} escape fraction of EoR galaxies?

This paper is organized as follows. Section \ref{sec:data} describes the NIRCam and NIRSpec observations used in this study. In Section \ref{sec:analysis}, we describe how we measure the {\lya} line properties of the targeted galaxies, and present evidence of less-attenuated {\lya} lines from galaxies near the luminous quasars. In Section \ref{sec:discussion}, we evaluate the escape fraction of galaxies near quasars utilizing the high IGM transmission around quasars, and discuss the implications for reionization models. We make conclusions in Section \ref{sec:conclusion}.
Throughout this paper, we adopt a flat $\Lambda$CDM cosmology with $H_0=70\text{ km s}^{-1}$ and $\Omega_M=0.3$, and use the AB magnitude system \citep{ABsystem}.

\section{Data} \label{sec:data}

\subsection{NIRCam Observations}

This work focuses on galaxies observed in two high-redshift quasar fields, namely around the quasars J0100+2802 $(z=6.327)$ and J1148+5251 $(z=6.42)$.
These quasar fields were observed as part of the {\em Emission-line galaxies and Intergalactic Gas in the Epoch of Reionization} (EIGER) program (GO-1243, PI: Lilly), which delivers NIRCam F115W, F200W, and F356W imaging and F356W grism spectroscopy.
Figure \ref{fig:field} shows the NIRCam images of the two quasar fields.
The images have sizes of $3'\times6'$, which corresponds to $\approx 1\times2$ proper Mpc at the quasars' redshifts.
We refer the readers to \citet{kashino23} and \citet{matthee23} for details on the data reduction of the EIGER observations. 
The typical $5\sigma$ depths of the F115W, F200W, and F356W imaging are 27.8, 28.3, and 28.1 magnitudes, respectively.
The sensitivity of the F356W grism spectra depends on the wavelength and the source positions, with the best
sensitivity reaching $0.6\times10^{-18}\text{ erg s}^{-1}\text{ cm}^{-2}$ at 3.8 micron. 

Using the F356W grism spectra,
\citet{matthee23} and \citet{kashino25} identified {\oiii}-emitters in the two quasar fields and performed spectral energy distribution (SED) fitting for these galaxies. 
In this work, we use the {\oiii} redshifts as the systemic redshifts $(z_\text{sys})$ of galaxies.
We also estimate the absolute magnitude at rest-frame $1500\text{\AA}$ $(M_\text{UV})$ of all the galaxies by fitting a power law SED to the F115W and F200W magnitudes. These galaxy properties can be found in Table \ref{tbl:sample}. 

\subsection{NIRSpec/MSA Spectroscopy}

We obtain the NIRSpec/MSA G140M/F070LP spectroscopy for the two quasar fields as part of the {\em Mapping Super-luminous Quasars’ Extended Radiative Emission} (MASQUERADE) project (GO-3117 and GO-4713, PI: Eilers). 
The spectra cover wavelengths from $8000\text{\AA}\lesssim\lambda_\text{obs}\lesssim13000\text{\AA}$, 
with a wavelength-dependent spectral resolution of $R\approx600-1000$. At the quasars' {\lya} wavelengths $(\lambda_\text{obs}\approx9000${\AA}), the spectral resolution is $R\approx600$. Each quasar field is covered by two MSA pointings, and we adopt the 3-Shutter dither pattern for the observations. The total on-source exposure time on each pointing is 7.7 hours. 

\begin{figure*}
    \centering
    \includegraphics[width=0.49\linewidth, trim={1.7cm 1cm 1.2cm 1.5cm}, clip]{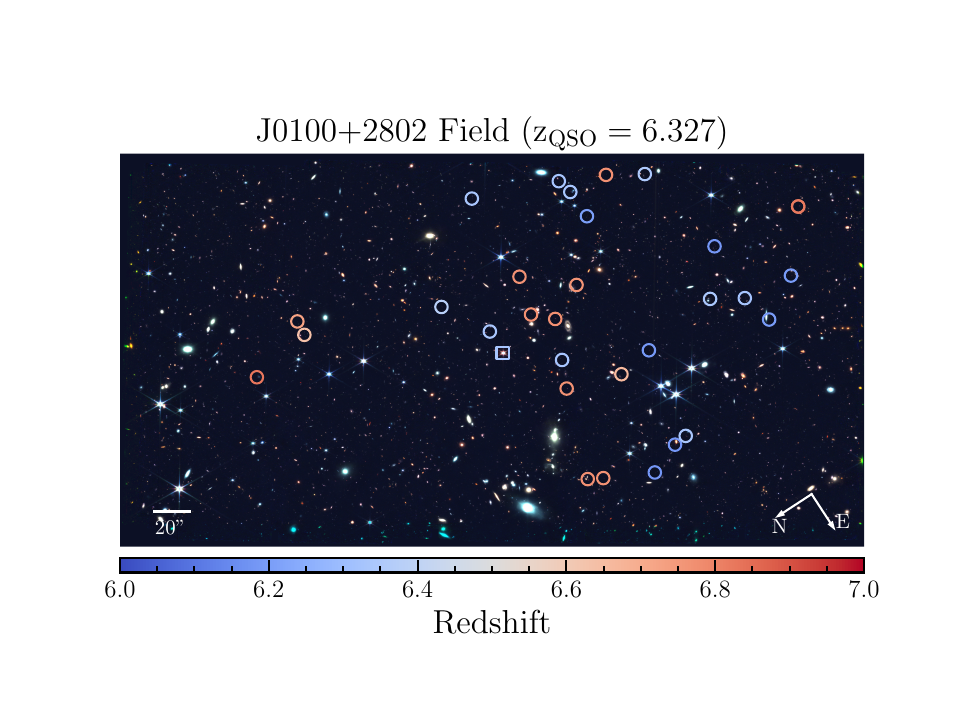}
    \includegraphics[width=0.49\linewidth, trim={1.7cm 1cm 1.2cm 1.5cm}, clip]{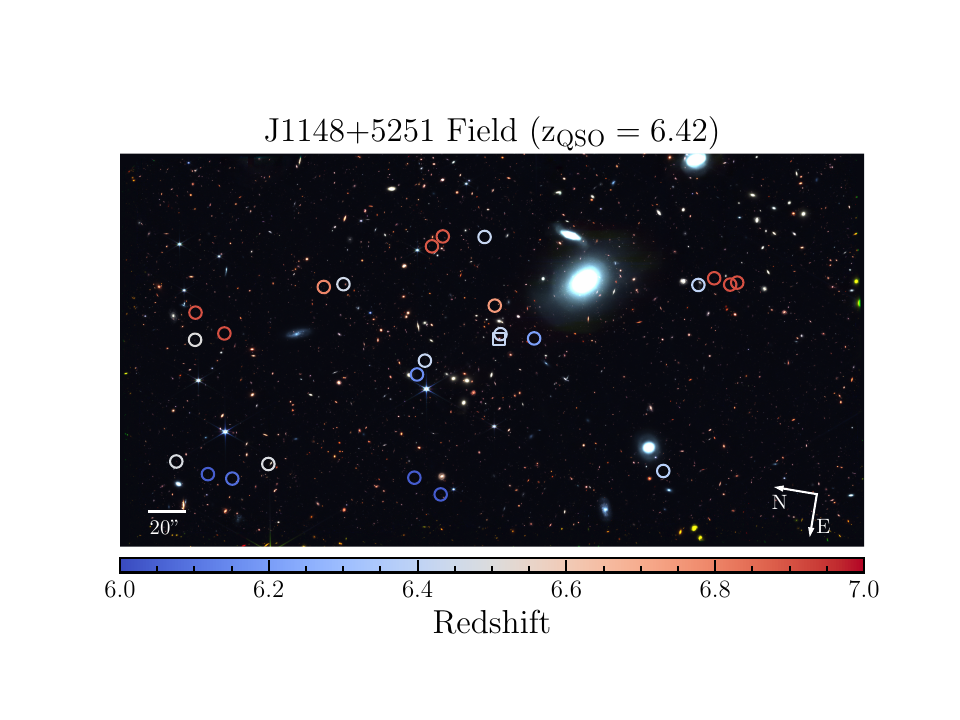}
    \caption{The two quasar fields targeted by this work, J0100+2802 (left) and J1148+5251 (right). We compose the RGB images using the NIRCam F115W, F200W, and F356W images from the EIGER project \citep[e.g.,][]{kashino23}. The quasars are at the field centers, marked by the squares. The circles mark the galaxy sample of this study, with the colors representing their systemic redshifts. There are 28 targets and 22 targets in the J0100+2802 and J1148+5251 fields, respectively.}
    \label{fig:field}
\end{figure*}

The primary targets of the MSA observation are {\oiii}-emitters at $z_\text{sys}>5.5$ in the quasar fields. In this study, we focus on galaxies at $z_\text{sys}>6$ and with $M_\text{UV} < -19$, as the MSA observations are not sufficiently deep to detect the continuum of fainter sources. The circles in Figure \ref{fig:field} show the location of the galaxies we analyze in this work, and the galaxy coordinates can be found in Table ~\ref{tbl:sample}.

We reduce the spectra by running the {\texttt{Detector1Pipeline}} using {\texttt{jwst}}\footnote{https://jwst-pipeline.readthedocs.io/en/stable/} version 1.17.1 and CRDS jwst\_1322.pmap to get the rate files, then using {\texttt{msaexp}}\footnote{https://github.com/gbrammer/msaexp/tree/main/msaexp} version 0.9.5.dev8+ge2b237b \citep{msaexp} to extract the 2D spectra. As the final step, we use PypeIt \citep{pypeit}\footnote{https://pypeit.readthedocs.io/en/stable/} version 1.17.1 to coadd the 2D spectra for individual objects and extract their 1D spectra.

{Although not analyzed in this work, we also extract the spectra of $M_\text{UV} > -19$ galaxies targeted by the MSA observation (36 galaxies in total). These spectra are available online\footnote{https://github.com/cosmicdawn-mit/MASQUERADE\_LAE/tree/main}. Detailed analysis for galaxies will be presented in future studies.}


\section{Unattenuated {\lya} Lines from Galaxies Near Quasars} \label{sec:analysis}

Luminous quasars at $z>6$ are expected to produce large ionized bubbles around them, where the IGM has low opacity to {\lya} photons.
In this section, we test whether we can measure the {\lya} lines from galaxies near quasars without significant IGM attenuation. 
We first fit the galaxy spectra to measure their {\lya} fluxes, then compare the {\lya} profiles and equivalent widths of galaxies near quasars to galaxies outside the quasars' ionized bubbles. 

\begin{deluxetable*}{c|cccccccc}
\label{tbl:sample}
\tablecaption{Properties of galaxies in our sample}
\tablewidth{0pt}
\tablehead{\colhead{ID} & \colhead{RA} & \colhead{Dec} &  \colhead{$z_\text{sys}$} &  \colhead{$M_\text{UV}$} & \colhead{$z_{\text{Ly}\alpha}$} & \colhead{${F}_{\text{Ly}\alpha}$} & \colhead{\ewlya} & \colhead{\fesclya}\\
\colhead{} & \colhead{(deg)} & \colhead{(deg)} & \colhead{} & \colhead{(mag)} & \colhead{} & \colhead{$(10^{-18}\text{erg s}^{-1}\text{cm}^{-2})$} & \colhead{({\AA})} & \colhead{}}
\startdata
\hline
J1148\_2637 & 177.112276 & 52.869027 & 6.037 & --20.45 &  & $<2.76$ & $<12.19$ & $<0.03$ \\
J1148\_3391 & 177.117208 & 52.902130 & 6.075 & --20.07 &  & $<0.59$ & $<2.48$ & $<0.02$ \\
J1148\_3461 & 177.109067 & 52.873590 & 6.039 & --20.24 &  & $<0.70$ & $<3.59$ & $<0.08$ \\
J1148\_3608 & 177.117109 & 52.906070 & 6.042 & --19.19 & $6.045^{+0.001}_{-0.001}$ & $3.10^{+0.55}_{-0.47}$ & $34.34^{+10.88}_{-8.04}$ & $0.30\pm{0.25}$ \\
J1148\_3822 & 177.096543 & 52.834714 & 6.375 & --19.97 &  & $<1.71$ & $<90.55$ & $<0.60$ \\
J1148\_4246 & 177.111852 & 52.896855 & 6.485 & --19.63 &  & $<2.50$ & $<27.07$ & $<0.27$ \\
\hline
\enddata
\tablecomments{The entire table will be available online upon the acceptance of this paper.}
\end{deluxetable*}

\subsection{Fitting the {\lya} Lines} \label{sec:analysis:lae}

We fit the {\lya} lines of galaxies targeted by the NIRSpec/MSA observation. We exclude five objects that have their {\lya} lines falling in the detector gap, and three objects that are severely contaminated by nearby bright sources. For each remaining object, we fit the spectral window $1200\text{\AA}<\lambda_\text{rest}<1500\text{\AA}$, where $\lambda_\text{rest}=\lambda_\text{obs}/(1+z_\text{sys})$ is the rest-frame wavelength. We use a power-law continuum plus a Gaussian profile to model the spectral region around the \lya\ emission, where the continuum flux at $\lambda_\text{rest}<\lambda_{\text{Ly}\alpha}$ is set to be zero. 
This model contains five free parameters, namely an amplitude $(A)$ and a slope $(\beta)$ of each galaxy's continuum, the {\lya} redshift $(z_{\text{Ly}\alpha})$, the {\lya} flux $(F_{\text{Ly}\alpha})$, and the line width $(\sigma_{\text{Ly}\alpha})$. We run a nested sampling algorithm {\texttt{dynesty}} to determine the posterior probability distribution for each model parameter, where we adopt flat priors for all the parameters. {We adopt the median of the posterior samples as fitting results,}
and determine their uncertainties from the $16$th and $84$th percentile. 
{We test the impact of priors by running the fit with Gaussian priors, and find that the specific choice for priors has negligible impact on the fitting result.}

Given the spectral resolution of the G140M grating, which is approximately $500\text{ km s}^{-1}$ at the {\lya} wavelength, the {\lya} lines are mostly unresolved or only marginally resolved. Therefore, we use a Gaussian profile to describe the {\lya} line without modeling more complex line structures (such as asymmetries and outflows). This approach is sufficient for measuring the {\lya} line fluxes and {equivalent widths}, but does not allow us to identify any double-peak \lya\ emitters, as the peak separation is typically $\lesssim500~\rm km\,s^{-1}$ \citep[e.g.,][]{verhamme18,naidu22}. 

Note that we aim to test whether galaxies near quasars show less attenuated {\lya} lines compared to galaxies outside the quasars' ionized bubble.
Therefore, we need to exclude possible AGNs from our sample, as AGNs may also produce their own ionized bubbles and show unattenuated {\lya} lines.
We identify possible AGNs by measuring the {\civ} line fluxes
 \citep[e.g.,][]{nakajima18,saxena22}. For each galaxy in our sample, we fit its spectrum at $1500\text{\AA}<\lambda_\text{rest}<1600\text{\AA}$ as a power-law continuum plus a Gaussian emission line.
 We find five galaxies with rest-frame {\civ} EW higher than 12{\AA}; according to \citet{nakajima18}, these objects are likely AGNs, and we exclude them from the rest of this Section. We present more information about these {\civ} emitters in the Appendix.


The final sample for this work consists of 50 galaxies at $6<z<7$ with $M_\text{UV}<-19$, including 28 galaxies in the J0100+2802 field and 22 galaxies in the J1148+5251 field. The median $M_\text{UV}$ of this sample is $-19.97$. Of these, 14 galaxies have a {\lya} line detection with $>3\sigma$ significance. Table \ref{tbl:sample} summarizes the properties of the galaxies in our sample. 

{
We end this subsection by discussing the faint galaxies $(M_\text{UV}<-19)$ targeted by our MSA observation, which are not included in the final sample. We fit the {\lya} lines for these galaxies, and 12 of them have {\lya} lines detections at $>3\sigma$ levels. The {\lya} line fluxes and equivalent widths (or the upper and lower limits of these quantities) for these $M_\text{UV}<-19$ galaxies are available online at the GitHub repository\footnote{https://github.com/cosmicdawn-mit/MASQUERADE\_LAE/tree/main} for the reader's reference.}


\subsection{The {\lya} Profile of Galaxies Near Quasars} \label{sec:analysis:stack}

We now investigate the {\lya} line profiles for galaxies near the quasars.
Due to the IGM's absorption, {most $z>6$ galaxies have no flux immediately blueward of their systemic {\lya} wavelengths.}
In contrast, for galaxies residing within ionized bubbles generated by luminous quasars, the blue wing of their {\lya} lines is expected to be visible due to the low IGM opacity. 
To investigate this effect, we stack the spectra of galaxies in our sample, as
 the spectra of individual galaxies do not have sufficient signal-to-noise ratio (S/N).
Specifically, we consider three subsets of the galaxies:
\begin{enumerate}
    \item $|z-z_{\rm Q}|<\frac{(1+z_{\rm Q})\Delta v_\text{max}}{c}$, where we take $\Delta v_\text{max}=2500\text{ km s}^{-1}$ in this work, and $z_Q$ is the quasar's redshift. This sample is referred to as ``galaxies near quasars";
    \item $6<z<z_{\rm Q}-\frac{(1+z_{\rm Q})\Delta v_\text{max}}{c}$, referred to as ``foreground galaxies";
    \item $z_{\rm Q}+\frac{(1+z_{\rm Q})\Delta v_\text{max}}{c}<z<7$, referred to as ``background galaxies".
\end{enumerate}
We note here that all the targeted galaxies have transverse distances to the quasars of $r_\perp\lesssim1$ proper Mpc. The expected sizes of the ionized bubble around $z>6$ luminous quasars is $\gtrsim1$ Mpc,
given the typical quasar lifetimes \citep[$\gtrsim10^6$ yr; e.g., ][]{eilers17, eilers25}. Therefore, we do not apply a cut on $r_\perp$ when selecting the ``galaxies near quasars" sample.


For each subset,
we select galaxies with at least $2\sigma$ detection for the {\lya} flux. We subtract the best-fit continuum model (from Section \ref{sec:analysis:lae}) from the spectrum of each galaxy and shift the spectrum to the rest-frame according to its systemic redshift. We then regrid the spectra to a common wavelength grid and stack the spectra using inverse-variance weighting. The uncertainties of the stacked spectra are derived by propagating the errors of the individual objects' spectra.

\begin{figure*}
    \centering
    \includegraphics[width=1\linewidth]{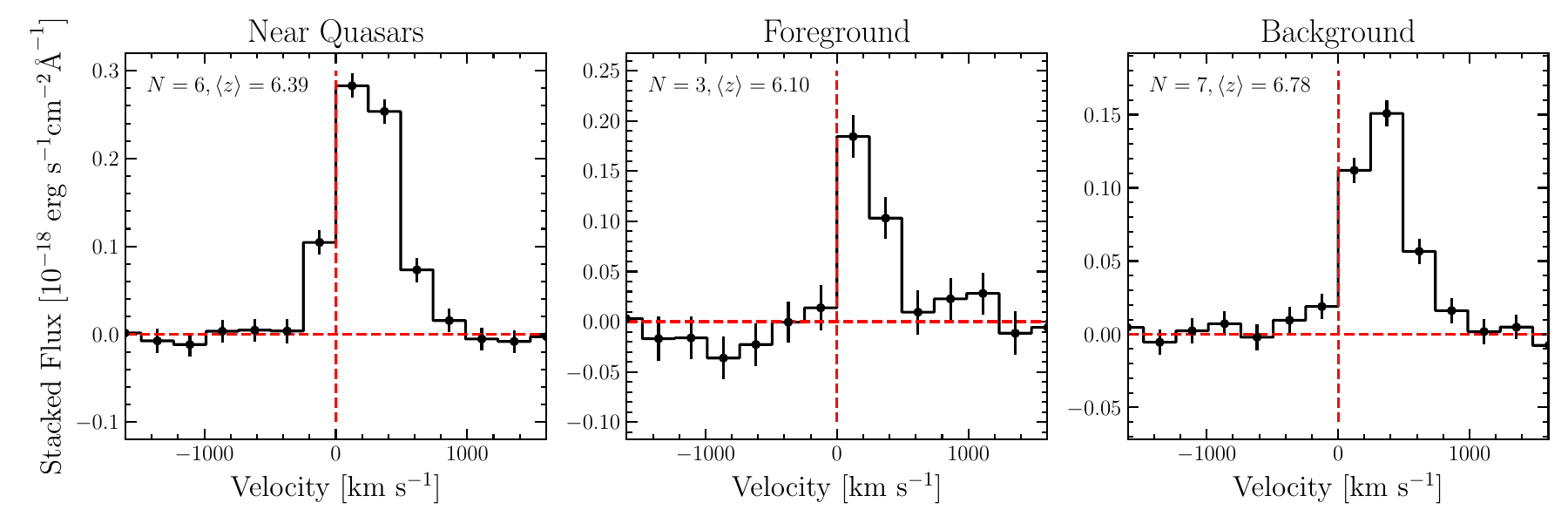}
    \caption{The stacked {\lya} line profile for galaxies near quasars (left), foreground galaxies (middle), and background galaxies (right). The text in each panel marks the number of galaxies and the average redshift for the stacked spectra. The dashed lines mark the systemic {\lya} wavelength and the level of zero flux. Galaxies near quasars show significant flux bluer than the systemic {\lya} wavelength, indicating low IGM opacity around quasars. In contrast, the foreground and background galaxies do not have significant flux bluer than the {\lya} wavelength.}
    \label{fig:spec}
\end{figure*}

Figure \ref{fig:spec} presents the stacked spectra. The stacked spectra of galaxies near quasars show significant flux bluer than the systemic {\lya} wavelength. In contrast, the stacked spectra for foreground and background galaxies do not have significant flux blueward of the systemic {\lya} wavelength. 
This result supports the scenario where quasars enhance the {\lya} transmission of their surrounding IGM, enabling the detection of {\lya} blue wings for galaxies near quasars. Again, we note that the G140M grating does not have sufficient resolution to unveil the complex line structures (e.g., the double peaks) of the {\lya} lines.

{To further validate our result and ensure the comparison in Figure \ref{fig:spec} is not dominated by individual objects, we bootstrap 100 times for the spectral stacking, and evaluate the mean and standard deviation of the bootstrapped stacked spectra. The bootstrapped spectra confirms our finding, i.e., the galaxies around quasars sample has strong a flux bluer than the {\lya} wavelength, while the other two samples have zero flux bluer than {\lya}.}


\subsection{The Ly$\alpha$ Equivalent Width Distribution} \label{sec:analysis:ew}
 

We then evaluate 
the rest-frame {\lya} equivalent width ({\ewlya}) of the targeted galaxies. 
If the IGM around quasars has low opacity to {\lya} photons, galaxies near high-redshift quasars should show higher {\ewlya} compared to other galaxies at similar redshifts. 

\begin{figure*}
    \centering
    \includegraphics[width=0.8\linewidth]{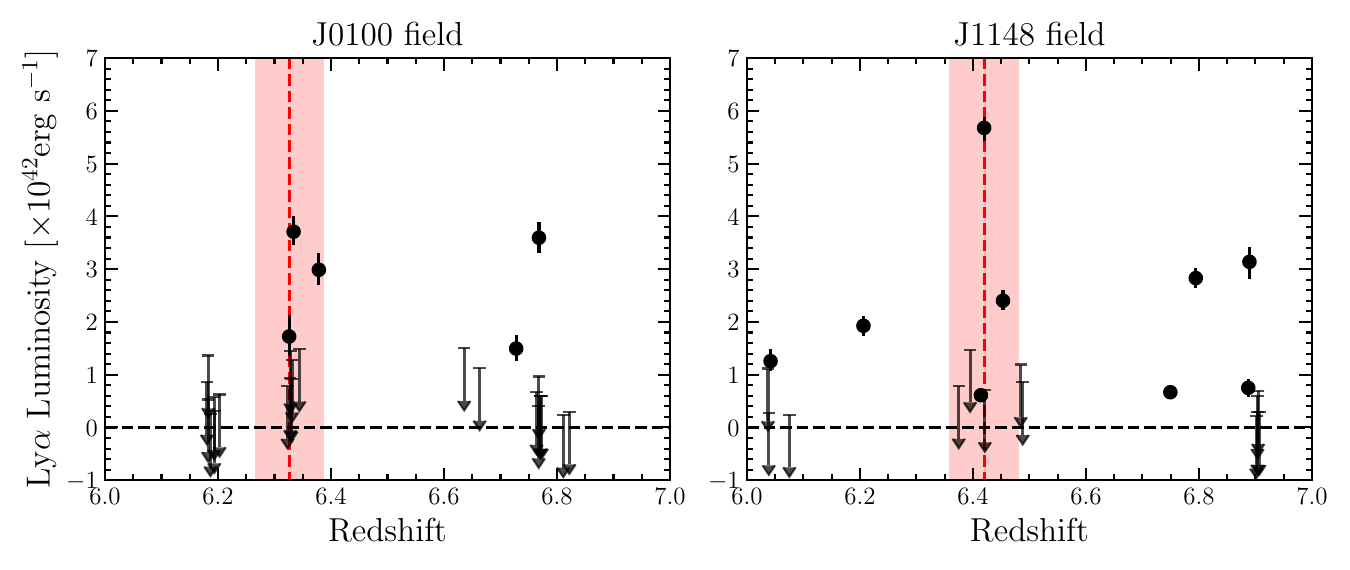}
    \includegraphics[width=0.8\linewidth]{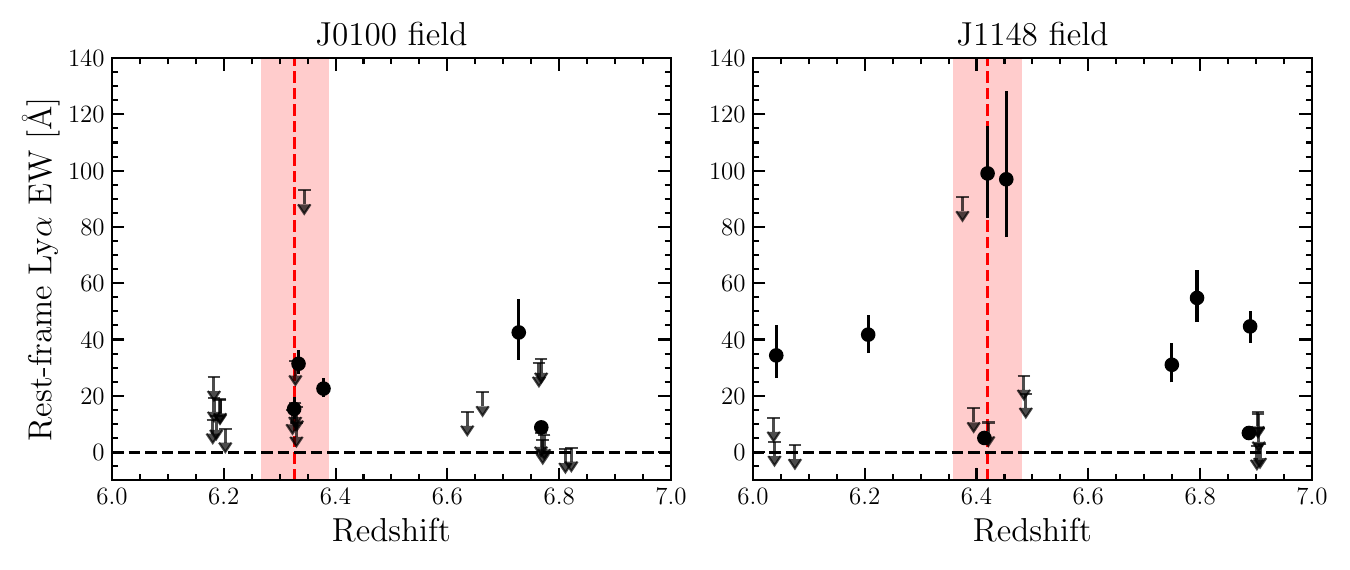}
    \includegraphics[width=0.8\linewidth]{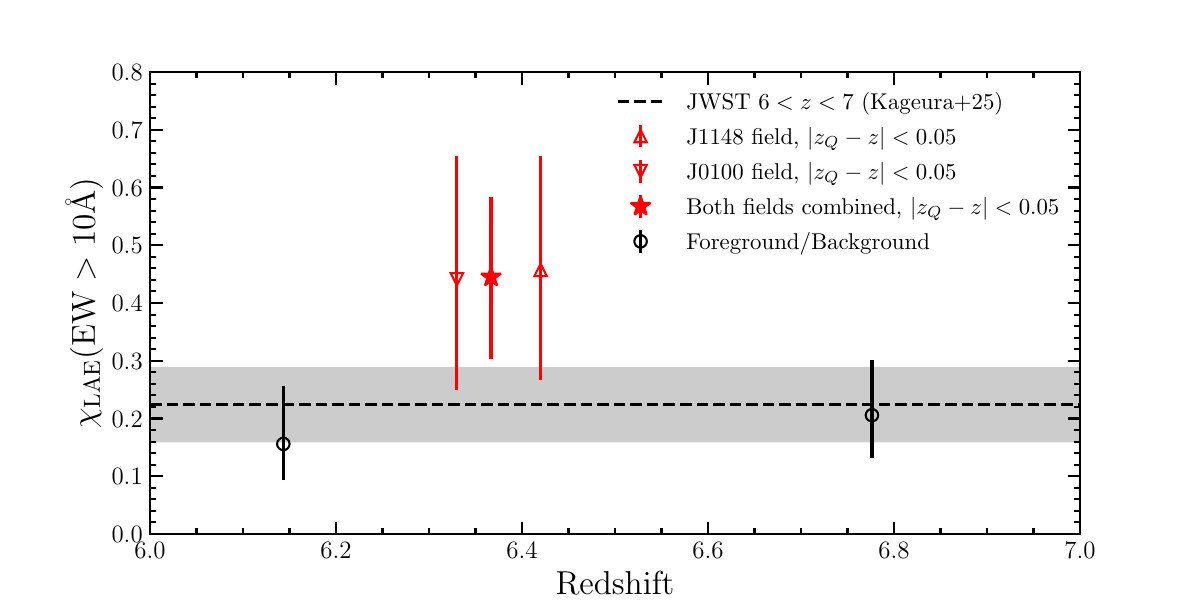}
    \caption{
    {\em Top Panels:} {redshift and {\lya} luminosity distribution of galaxies in the J0100 field and the J1148 field, in the left and right panel, respectively. The red shaded region marks the quasar's vicinity $|\Delta v|<2500\text{ km s}^{-1}$.
    {\em Middle Panels:}} Similar to the top panel, but for the redshift and {\lya} EW distribution. 
    {\em Bottom:} The fraction of galaxies with {\ewlya}$>10\text{\AA}$. This fraction is estimated by fitting the galaxy {\ewlya} with a log-normal distribution (see text for details). We also include the NIRSpec-observed galaxies at $6<z<7$ from \citet{kageura25} for comparison. In both quasar fields, galaxies in the quasar's vicinity show higher LAE fractions than foreground and background galaxies and the general galaxy population. This comparison indicates that IGM around luminous quasars has high transmission to {\lya} photons.}
    \label{fig:LyaEW}
\end{figure*}

The top {and middle} panel of Figure \ref{fig:LyaEW} shows {the {\lya} luminosity and equivalent width } 
of individual galaxies in our sample. The red dashed line marks the quasars' redshifts, and the shaded region represents the region around the quasars (i.e., $\Delta v<2500\text{ km s}^{-1}$). 
Among the 15 galaxies near quasars, six $(40\%)$ of them have {\lya} lines detected at higher than $3\sigma$ significance. In contrast, this fraction is only $8/35=22.9\%$ for foreground and background galaxies. 
The fractions of galaxies with {\ewlya}$>10\text{\AA}$ are $5/15=33.3\%$ and $7/35=20.0\%$ for galaxies near quasars and foreground/background galaxies, respectively.
This comparison already hints that galaxies around quasars have less attenuated {\lya} emissions, due to the low IGM opacity in the quasar's proximity zone.


To further quantify the difference in {\lya} emissions between galaxies inside and outside the quasars' ionized bubbles, we 
 estimate the distribution of {\ewlya} for galaxies in our sample. {We follow the Bayesian analysis in \citet{tang24a} \citep[also see][]{schenker14, endsley21, boyett22, chen24}. Specifically, } we assume a log-normal distribution for {\ewlya}:

\begin{equation} \label{eq:1}
    p(x|\mu,\sigma) = \frac{A}{\sqrt{2\pi}\sigma x}\times \exp\left[-\frac{(\ln x-\mu)^2}{2\sigma^2}\right]
\end{equation}
where $x$ is the random variable ({\ewlya} in this case), $\mu$ and $\sigma$ are the mean and standard deviation of the log-normal distribution, and $A$ is the normalization. We derive the posterior distribution of $(\mu, \sigma)$ according to Bayes' theorem:
\begin{equation}
    p(\mu,\sigma|\text{obs})\propto p(\text{obs}|\mu,\sigma)p(\mu,\sigma)
\end{equation}
where $p(\mu,\sigma)$ is the prior, and we adopt flat priors for $\mu$ and $\sigma$. $p(\text{obs}|\mu,\sigma)$ is the probability of getting the observed {\ewlya} for the galaxies. Specifically,
\begin{multline}
    p(\text{obs}|\mu,\sigma)=\prod_i p(\text{obs}, i|\mu,\sigma)\\
    =\prod_i \int_0^{+\infty} p(x)_{\text{obs},i} \cdot p(x|\mu,\sigma) \diff x
\end{multline}
where $i$ goes through all galaxies in the sample, and $p(x)_{\text{obs},i}$ is the posterior distribution of the galaxies' {\ewlya}. For galaxies with $>3\sigma$ {\lya} line detection, we assume a Gaussian error for the detection, i.e.,
\begin{equation}
    p(x)_{\text{obs},i} = \frac{1}{\sqrt{2\pi}\sigma_{\text{obs},i}}\cdot \exp\left[-\frac{(x-x_{\text{obs},i})^2}{2\sigma_{\text{obs},i}^2}\right]
\end{equation}
For upper limits, we have
\begin{equation} \label{eq:last}
    p(\text{obs}, i|\mu,\sigma) = p(x<x^{\text{lim}}_{3\sigma,i}|\mu,\sigma)
\end{equation}
We run Markov Chain Monte Carlo (MCMC) to sample the posterior of $(\mu, \sigma)$.

Using this method, we model the {\ewlya} distributions for galaxies near quasars, foreground galaxies, and background galaxies.
For comparison, we also fit the distributions for {NIRSpec}-observed $6<z<7$ galaxies compiled by \citet{kageura25}. When fitting the galaxies from \citet{kageura25}, we apply a cut of $M_\text{UV}<-19$ to match the absolute magnitude of our sample.

Table \ref{tbl:distributions} lists the best-fit parameters for these galaxy subsets. 
Using the posterior of $(\mu,\sigma)$, we also compute the {\lya}-emitter (LAE) fractions (defined as galaxies with {\ewlya}$>10\text{\AA}$) for the subsets of galaxies, which are shown in the bottom panel of Figure \ref{fig:LyaEW}. The foreground and background galaxy samples have LAE fractions close to the $6<z<7$ galaxy sample from \citet{kageura25}. In contrast, in both quasar fields, galaxies near quasars show higher LAE fractions. This comparison again indicates that galaxies near quasars have less-attenuated {\lya} lines compared to other $z>6$ galaxies, due to the low IGM opacity around quasars.
Note that the UV slope and UV magnitude distributions of galaxies near quasars are very similar to those of foreground and background galaxies; therefore, the difference in {\ewlya} can only be attributed to the difference in IGM opacity, i.e., the influence of the quasars' ionizing radiation fields.

We notice that galaxies near quasars have LAE fractions $\mu(\text{EW}_{\text{Ly}\alpha}^{\text{rest}})$ 
similar to $z\sim5$ galaxies measured by \citet{tang24a}. Our result indicates that the unattenuated {\lya} line EW of $M_\text{UV}<-19$ galaxies has little evolution from $z\sim6.4$ to $z\sim5$.

{We also} note that including {\civ} emitters into the {\ewlya} distribution has a negligible impact on the derived LAE fractions. By including the {\civ} emitters, the {\ewlya}$>10\text{\AA}$ fraction of galaxies near quasars changes to $\chi_\text{LAE}(\text{EW}>10\text{\AA})=0.44_{-0.14}^{+0.15}$, and that of foreground + background galaxies changes to $\chi_\text{LAE}(\text{EW}>10\text{\AA})=0.24_{-0.06}^{+0.07}$.

\begin{deluxetable*}{c|cc|cc|cc}
\label{tbl:distributions}
\tablecaption{The {\ewlya} distribution of galaxies} \label{tbl:dist}
\tablewidth{0\columnwidth} 
\tablehead{\colhead{Sample} & \colhead{$\langle z \rangle$} & \colhead{\makecell{Detection \\ Fraction \tablenotemark{1}}} &
\colhead{$\mu(\text{EW}_{\text{Ly}\alpha}^{\text{rest}})$} & \colhead{$\sigma(\text{EW}_{\text{Ly}\alpha}^{\text{rest}})$} & 
\colhead{$\chi_\text{LAE}(\text{EW}>10\text{\AA})$} &
\colhead{$\chi_\text{LAE}(\text{EW}>25\text{\AA})$}\\
\colhead{} & \colhead{} & \colhead{} &
\colhead{({\AA})} & \colhead{({\AA})} & 
\colhead{} &
\colhead{}
}
\startdata
\hline
Galaxies near Quasars & 6.37 & 6/15 &  $2.07_{-0.89}^{+0.53}$ & $1.70_{-0.51}^{+0.89}$ & $0.45_{-0.14}^{+0.14}$ & $0.24_{-0.08}^{+0.11}$ \\
Foreground Galaxies & 6.14 & 2/12 & $-0.89_{-1.40}^{+1.63}$ & $3.14_{-1.08}^{+1.10}$ & $0.16_{-0.06}^{+0.10}$ & $0.09_{-0.05}^{+0.07}$ \\
Background Galaxies & 6.71 & 6/23 & $0.03_{-1.23}^{+0.91}$ & $2.86_{-0.78}^{+1.05}$ & $0.21_{-0.07}^{+0.10}$ & $0.12_{-0.05}^{+0.08}$ \\
Foreground + Background & 6.53 & 8/35 & $0.00_{-1.20}^{+0.78}$ & $2.61_{-0.68}^{+0.99}$ & $0.19_{-0.06}^{+0.06}$ & $0.10_{-0.04}^{+0.05}$ \\\hline
\citet{kageura25} \tablenotemark{2} & 6.42 & -- & $0.76_{-0.92}^{+0.63}$ & $1.96_{-0.46}^{+0.76}$ & $0.22_{-0.06}^{+0.07}$ & $0.11_{-0.04}^{+0.04}$ \\\hline
\enddata
\tablenotetext{1}{{The fraction of galaxies with at least $3\sigma$ detection of their {\lya} line.}}
\tablenotetext{2}{Galaxies observed by NIRSpec at $6<z<7$, as compiled by \citet{kageura25}.}
\tablecomments{All samples are selected to have $M_\text{UV}<-19$.}
\end{deluxetable*}

\subsubsection{How do the Quasars' Radiation Affect Galaxy Properties?}


{
The analysis above assumes that the quasars' radiation fields do not significantly affect the intrinsic {\lya} properties of their surrounding galaxies. If the quasars' radiation enhances the star formation and/or the {\lya} escape fraction of nearby galaxies, we expect to see a high LAE fraction for galaxies near quasars even without a low IGM opacity. In this subsection, we discuss the possible impact of the quasars' radiation field on galaxy properties.}

{
We first evaluate the strength of a quasar's radiation field at the position of its surrounding galaxies. The brighter quasar in our sample, J0100+2802, has $M_\text{UV}=29$ \citep{Mazzucchelli23}. The faintest galaxy in our sample has $M_\text{UV}<-19$, i.e., $10^4$ times fainter than the quasar. The transverse distances from the galaxies to the quasars are $r_\perp\gtrsim0.5$ Mpc. Assuming a typical galaxy size of $1$ kpc, we estimate the quasars' radiation fields should be $\lesssim4\%$ of the galaxies' own radiation. This comparison suggests that quasars should have little impact on the surrounding galaxies' star formation. This argument is strengthened by the short lifetimes of quasars \citep[$\lesssim1$ Myr, e.g.,][]{Khrykin21,eilers25}, given that the {\lya} line traces the star formation in the past several Myrs \citep[e.g.,][]{sobral19}.}

{
Another way to investigate the potential impact of quasars on their surrounding galaxies is by artificially adding IGM attenuation to the galaxies near quasars. The IGM at $z\sim6.3-6.4$ primarily absorbs the blue wings of {\lya} lines without causing significant damping absorption for the red wings; therefore, we expect that the increased transmission in the vicinity of quasars enhances the {\lya} flux of galaxies near quasars by no more than a factor of two. If we reduce the observed {\lya} flux of galaxies near quasars by a factor of two and still see an enhancement in {\lya} emission, it will indicate that quasar radiation boosts the {\lya} emission in nearby galaxies.}

{Following this argument, we divide the {\lya} fluxes of the galaxies near quasars by two and re-perform the analysis. The derived LAE fraction for galaxies near quasars is $0.28^{+0.11}_{-0.09}$, close to the $6<z<7$ galaxy sample. Figure \ref{fig:LyaEW} also implies that, with their {\lya} fluxes divided by two, galaxies near quasars will be indistinguishable from foreground and background galaxies. In other words, our dataset shows no clear evidence that quasar radiation significantly boosts the intrinsic {\lya} emission of galaxies near quasars.}

{Finally, we note that our dataset is not able to fully distinguish the influence of IGM opacity and intrinsic {\lya} line properties for galaxies near quasars. Future high-resolution spectroscopy for these galaxies will provide resolved {\lya} profiles of these galaxies, which will enable modeling of the IGM transmission profile around the quasars and break the degeneracy discussed above.}

\section{Implication for the Escape Fraction of Reionization-Epoch Galaxies}  \label{sec:discussion}

The enhanced {\ewlya} and the detection of the {\lya} blue wing for galaxies near quasars indicate that we can detect the {\lya} emission line from these galaxies without significant IGM attenuation. These galaxies provide a rare opportunity to directly constrain the {\lya} emission lines properties of $z>6$ galaxies, which is usually impossible due to the high IGM neutral fraction \citep[$x_\text{HI}\gtrsim0.1$, e.g.,][]{jin23, durovcikova24}. 
As an example, in this section, we use galaxies near quasars in our sample to put direct constraints on the {\lya} escape fraction ({\fesclya}) of $z>6$ galaxies.

To evaluate {\fesclya}, for each galaxy, we adopt the {\hb} flux from the {\oiii} emitter catalog presented in \citet{matthee23} and \citet{kashino25}. We note that one galaxy does not have {\hb} measurement and is excluded from the escape fraction analysis. We obtain the dust attenuation $(A_V)$ from the catalog and correct the {\hb} fluxes for dust attenuation. The median attenuation of the sample is $0.22$ {mag}. We also estimate the continuum flux at $\lambda_\text{rest}=1215.67\text{\AA}$ (denoted by $F_\text{cont}^{1215.67\text{\AA}}$) by fitting its F115W and F200W magnitudes using a power law SED. We then compute the escape fraction as
\begin{equation}
    f_\text{esc}^{\text{Ly}\alpha} = \frac{\text{EW}_{\text{Ly}\alpha}^\text{rest}\times F_\text{cont}^{1215.67\text{\AA}}}{25\times F_{\text{H}\beta}}
\end{equation}
where {\ewlya} comes from fitting the MSA spectra, and the coefficient 25 corresponds to case B recombination with temperature $T=10^4K$ \citep[e.g.,][]{tang24b,kageura25}.
This approach avoids possible impacts due to the slitloss of the MSA slitlets.

With the {\fesclya} of individual galaxies, we compute the distribution of {\fesclya} following the same method we use for evaluating the {\ewlya} distribution as described in Section \ref{sec:analysis:ew}. 
Specifically, we assume a lognormal distribution for {\fesclya}, and use Equations \ref{eq:1} to \ref{eq:last} to evaluate the posterior of $(\mu,\sigma)$ for the distribution.
For galaxies near quasars, we get $\mu=-2.78^{+1.48}_{-1.19}$ and $\sigma=3.41^{+1.06}_{-1.39}$. These numbers indicate an average escape fraction of $f_\text{esc}^{\text{Ly}\alpha}=0.14\pm0.04$, consistent with the measurements for $z\lesssim5$ galaxies \citep[e.g.,][]{lin24}.

The LyC escape fraction is positively correlated with the {\lya} escape fraction, and most galaxies have $f_\text{esc}^\text{LyC}<f_\text{esc}^{\text{Ly}\alpha}$ \citep[e.g., ][]{dijkstra16,verhamme16}.
We estimate the average LyC escape fraction 
for galaxies near quasars using the relation in \citet{begley24}, who suggested that $f_\text{esc}^{\text{LyC}}=0.15\times f_\text{esc}^{\text{Ly}\alpha}$. This relation gives 
$f_\text{esc}^{\text{LyC}}\approx0.02\pm0.01$. We note that the $f_\text{esc}^{\text{LyC}}-f_\text{esc}^{\text{Ly}\alpha}$ relation has large scatter and might depend on the properties of galaxies; therefore, we take $f_\text{esc}^{\text{LyC}}<f_\text{esc}^{\text{Ly}\alpha}\approx0.14$ as a conservative upper limit. 
Due to the small sample size (15 galaxies in total, where 6 galaxies have {\lya} detections), we are not able to further explore the correlation between {\fesclya} and other galaxy properties \citep[like $M_\text{UV}$ and UV-slope; e.g., ][]{anderson17, Chisholm22}.

Our analysis demonstrates the unique power of high-redshift quasar fields in measuring the escape fraction of EoR galaxies. The estimated {\fesclyC} disfavors the ``high escape fraction" scenario with $f_\text{esc}^\text{LyC}\gtrsim0.2$ \citep[e.g.,][]{naidu20}, at least for galaxies with $M_\text{UV}<-19$ at $z\sim6.4$. 
Given the UV luminosity functions of $z>6$ galaxies \citep[e.g.,][]{bouwens22}, a low escape fraction of $f_\text{esc}^\text{LyC}\approx2\%$ indicates that luminous galaxies (e.g., those with $M_\text{UV}<-18$) cannot make a major contribution to the ionizing photon production.
In a recent study, \citet{simmons24} measured the ionizing photon production efficiency $(\xi_\text{ion})$ for $3<z<9$ galaxies using NIRCam imaging. With the $\xi_\text{ion}(M_\text{UV},z)$ they presented, we only need a low escape fraction $f_\text{esc}^\text{LyC}\lesssim0.05$ to reionize the universe. This picture is consistent with our measurement. The estimated {\fesclyC} is also consistent with the finding of \citet{papovich25}, who performed stellar population and nebular emission modeling for $4.5 < z < 9.0$ observed using {\em JWST} imaging and prism spectroscopy and found an average escape fraction of $f_\text{esc}^\text{LyC}=0.03\pm0.01$. 

\subsection{Systematic Uncertainties}

In order to measure the {\fesclya} for galaxies near quasars, we need to assume 
that these galaxies have negligible IGM attenuation for their {\lya} lines. Here we discuss the systemic uncertainties introduced by this assumption.  
The {\lya} blue wing in the stacked spectra (Figure \ref{fig:spec}) indicates that the IGM opacity around the quasars is similar to that of $z\sim5$ IGM \citep[e.g.,][]{hayes20}. 
This similarity is further supported by the recent transverse proximity effect measurement for J0100+2802 \citep{eilers25}. In short, \citet{eilers25} measured the transmission flux for background galaxies $(z>z_Q+0.1)$ and found a {\lya} transmission of $\mathcal{T}\gtrsim0.5$ for the IGM near the quasar. In comparison, IGM at $z\lesssim5$ has transmission of $\mathcal{T}\gtrsim0.1$ \citep[e.g.,][]{thomas21, meyer25}. 
We also notice that the IGM attenuation for {\lya} line fluxes is $\lesssim10\%$ at $z\lesssim5$ \citep[e.g.,][]{hayes20,matthee22}. 
We thus expect the uncertainty introduced by our assumption to be smaller than 10\%. This uncertainty does not influence our main result, i.e., the average {\lya} escape fraction is about $0.14$.

Another major uncertainty is from the limited sample size, as the ``galaxies near quasars" sample only contains 15 galaxies. Future NIRSpec/MSA observations for more high-redshift quasars will reduce the survey variance and provide tighter constraints on the escape fraction of $z>6$ galaxies.







\section{Conclusion}  \label{sec:conclusion}

We present {\em JWST} NIRSpec/MSA observations for galaxies in two high-redshift quasar fields, namely J0100+2802 $(z=6.327)$ and J1148+5251 $(z=6.42)$. We analyze 50 galaxies at $6<z<7$ with $M_\text{UV}<-19$, among which 15 are located near the quasars (with $\Delta v<2500\text{ km s}^{-1}$).
We measure the {\lya} fluxes, profiles, and rest-frame equivalent widths of these galaxies.
Leveraging the low IGM opacity around quasars, we are able to measure the {\lya} lines from galaxies near quasars without significant IGM attenuation.
This allows us to put direct constraints on the {\lya} escape fraction of $z>6$ galaxies.
Our main findings include

\begin{enumerate}
    \item The stacked {\lya} spectrum for galaxies near quasars clearly shows a significant emission bluer than the systemic {\lya} wavelength, 
    while foreground and background galaxies show no significant emission bluer than the systemic {\lya} wavelength in their stacked spectra. This result indicates that the {\lya} lines from galaxies near high-redshift quasars are less attenuated compared to other galaxies at similar redshifts.
    \item The LAE fraction of galaxies near quasars is $\chi_\text{LAE}(\text{EW}>10\text{\AA})=0.45_{-0.14}^{+0.14}$, higher than that of foreground and background galaxies ($0.19_{-0.06}^{+0.06}$). This result confirms that galaxies near quasars have less attenuated {\lya} lines.
    The {\ewlya} distribution of galaxies near quasars is similar to $z\sim5$ galaxies.

    \item 
    We estimate a median {\lya} escape fraction of {\fesclya}$=0.14\pm0.04$ for galaxies near quasars. Adopting the scaling relation of $f_\text{esc}^{\text{LyC}}=0.15\times f_\text{esc}^{\text{Ly}\alpha}$ from \citet{begley24}, we estimate $f_\text{esc}^{\text{LyC}}=0.02\pm0.01$ for galaxies near quasars. 
    This result favors the low escape fraction scenario $(f_\text{esc}^{\text{LyC}}\lesssim0.1)$ for $z>6$ galaxies.
    
\end{enumerate}

The ionized bubbles surrounding luminous quasars serve as unique laboratories for studying galaxy evolution and reionization in the early universe. High-resolution spectroscopy with NIRSpec's high-resolution grating or future extremely large telescopes will be able to resolve the {\lya} line profiles of galaxies near high-redshift quasars \citep[e.g.,][]{naidu22}. Such observations will enable key measurements for EoR galaxies, such as outflow properties and interstellar medium conditions. These experiments will offer valuable insights into the physical properties of EoR galaxies.

The {\it JWST} data used in this paper can be found in MAST: \dataset[10.17909/wm0b-5n06]{http://dx.doi.org/10.17909/wm0b-5n06}.
All the code and data used in this work will be available online at \url{https://github.com/cosmicdawn-mit/MASQUERADE_LAE/tree/main} upon the acceptance of this paper. 

\begin{acknowledgments}
{We thank the referee for valuable comments.}
This work is based on observations made with the NASA/ESA/CSA James Webb Space Telescope. The data were obtained from the Mikulski Archive for Space Telescopes at the Space Telescope Science Institute, which is operated by the Association of Universities for Research in Astronomy, Inc., under NASA contract NAS 5-03127 for JWST. These observations are associated with program ID $\#3117$ and $\#4713$. 
Support for this work was provided by NASA through the NASA Hubble Fellowship grant HST-HF2-51515.001-A awarded by the Space Telescope Science Institute, which is operated by the Association of Universities for Research in Astronomy, Incorporated, under NASA contract NAS5-26555.
\end{acknowledgments}

\vspace{5mm}
\facilities{JWST(NIRCam, NIRSpec)}


\software{PypeIt, msaexp, jwst, astropy\citep{astropy1,astropy2,astropy3}, bilby\citep{bilby_paper}}

\appendix

\section{{\civ} emitters in our sample}

In addition to the 50 galaxies in the main sample, we also identified five {\civ} emitters with $\text{EW}_0(\text{\civ})>12\text{\AA}$. These galaxies are possible AGNs. Three {\civ} emitters show {\lya} emissions at $>3\sigma$ levels, two of which have {\ewlya}$>100${\AA}. We highlight object J0100\_10287 at $z=6.77$, which has {\ewlya}$=152^{+60}_{-30}${\AA} and is the strongest {\lya} emitter in the MASQUERADE sample with $M_\text{UV}<-19$.
The properties of the {\civ} emitters are also included in Table \ref{tbl:sample}.

Following the method in Section \ref{sec:analysis:stack}, we produce the stacked {\lya} profile for the {\civ} emitters at the quasars' foreground and background. Although these galaxies are outside the quasars' ionized bubbles, the stacked {\lya} spectrum shows a significant {\lya} blue wing. This result indicates that these {\civ} emitters can produce their own ionized bubbles. Our finding is consistent with previous studies that suggest {\civ} emitters might be strong {\lya} leakers \citep[e.g.,][]{naidu22,Schaerer22,mascia23}.

We also note that only one {\civ} emitter is located in the ionized bubbles around quasars. Including this object in the ``galaxies near quasars" sample has a negligible impact on the estimated {\ewlya} and {\fesclya} distributions.

\begin{figure*}
    \centering
    \includegraphics[width=0.4\linewidth]{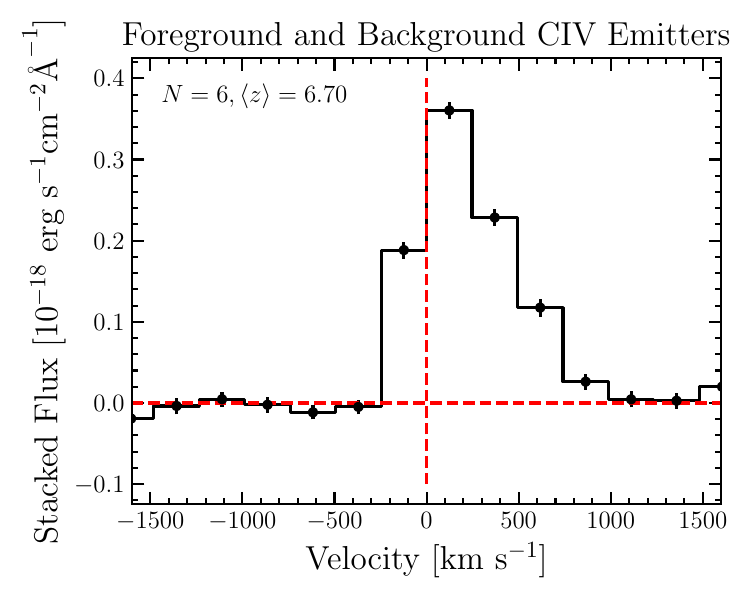}
    \caption{The stacked {\lya} line for {\civ} emitters at the quasars' foreground and background, showing significant flux bluer than the systemic {\lya} wavelength. This result indicate that {\civ} emitters can produce their own ionized bubbles.}
    \label{fig:civ_lya}
\end{figure*}

\bibliography{sample7}{}

\begin{thebibliography}{}
\expandafter\ifx\csname natexlab\endcsname\relax\def\natexlab#1{#1}\fi
\providecommand{\url}[1]{\href{#1}{#1}}
\providecommand{\dodoi}[1]{doi:~\href{http://doi.org/#1}{\nolinkurl{#1}}}
\providecommand{\doeprint}[1]{\href{http://ascl.net/#1}{\nolinkurl{http://ascl.net/#1}}}
\providecommand{\doarXiv}[1]{\href{https://arxiv.org/abs/#1}{\nolinkurl{https://arxiv.org/abs/#1}}}

\bibitem[{L. {Anderson} {et~al.}(2017){Anderson}, {Governato}, {Karcher}, {Quinn}, \& {Wadsley}}]{anderson17}
{Anderson}, L., {Governato}, F., {Karcher}, M., {Quinn}, T., \& {Wadsley}, J. 2017, \bibinfo{title}{{The little Galaxies that could (reionize the universe): predicting faint end slopes \& escape fractions at z>4},} \mnras, 468, 4077, \dodoi{10.1093/mnras/stx709}

\bibitem[{G. Ashton {et~al.}(2019)Ashton {et~al.}}]{bilby_paper}
Ashton, G., {et~al.} 2019, \bibinfo{title}{{BILBY: A user-friendly Bayesian inference library for gravitational-wave astronomy},} Astrophys. J. Suppl., 241, 27, \dodoi{10.3847/1538-4365/ab06fc}

\bibitem[{ {Astropy Collaboration} {et~al.}(2013){Astropy Collaboration}, {Robitaille}, {Tollerud}, {Greenfield}, {Droettboom}, {Bray}, {Aldcroft}, {Davis}, {Ginsburg}, {Price-Whelan}, {Kerzendorf}, {Conley}, {Crighton}, {Barbary}, {Muna}, {Ferguson}, {Grollier}, {Parikh}, {Nair}, {Unther}, {Deil}, {Woillez}, {Conseil}, {Kramer}, {Turner}, {Singer}, {Fox}, {Weaver}, {Zabalza}, {Edwards}, {Azalee Bostroem}, {Burke}, {Casey}, {Crawford}, {Dencheva}, {Ely}, {Jenness}, {Labrie}, {Lim}, {Pierfederici}, {Pontzen}, {Ptak}, {Refsdal}, {Servillat}, \& {Streicher}}]{astropy1}
{Astropy Collaboration}, {Robitaille}, T.~P., {Tollerud}, E.~J., {et~al.} 2013, \bibinfo{title}{{Astropy: A community Python package for astronomy},} \aap, 558, A33, \dodoi{10.1051/0004-6361/201322068}

\bibitem[{ {Astropy Collaboration} {et~al.}(2018){Astropy Collaboration}, {Price-Whelan}, {Sip{\H{o}}cz}, {G{\"u}nther}, {Lim}, {Crawford}, {Conseil}, {Shupe}, {Craig}, {Dencheva}, {Ginsburg}, {VanderPlas}, {Bradley}, {P{\'e}rez-Su{\'a}rez}, {de Val-Borro}, {Aldcroft}, {Cruz}, {Robitaille}, {Tollerud}, {Ardelean}, {Babej}, {Bach}, {Bachetti}, {Bakanov}, {Bamford}, {Barentsen}, {Barmby}, {Baumbach}, {Berry}, {Biscani}, {Boquien}, {Bostroem}, {Bouma}, {Brammer}, {Bray}, {Breytenbach}, {Buddelmeijer}, {Burke}, {Calderone}, {Cano Rodr{\'\i}guez}, {Cara}, {Cardoso}, {Cheedella}, {Copin}, {Corrales}, {Crichton}, {D'Avella}, {Deil}, {Depagne}, {Dietrich}, {Donath}, {Droettboom}, {Earl}, {Erben}, {Fabbro}, {Ferreira}, {Finethy}, {Fox}, {Garrison}, {Gibbons}, {Goldstein}, {Gommers}, {Greco}, {Greenfield}, {Groener}, {Grollier}, {Hagen}, {Hirst}, {Homeier}, {Horton}, {Hosseinzadeh}, {Hu}, {Hunkeler}, {Ivezi{\'c}}, {Jain}, {Jenness}, {Kanarek}, {Kendrew}, {Kern}, {Kerzendorf}, {Khvalko}, {King}, {Kirkby}, {Kulkarni},
  {Kumar}, {Lee}, {Lenz}, {Littlefair}, {Ma}, {Macleod}, {Mastropietro}, {McCully}, {Montagnac}, {Morris}, {Mueller}, {Mumford}, {Muna}, {Murphy}, {Nelson}, {Nguyen}, {Ninan}, {N{\"o}the}, {Ogaz}, {Oh}, {Parejko}, {Parley}, {Pascual}, {Patil}, {Patil}, {Plunkett}, {Prochaska}, {Rastogi}, {Reddy Janga}, {Sabater}, {Sakurikar}, {Seifert}, {Sherbert}, {Sherwood-Taylor}, {Shih}, {Sick}, {Silbiger}, {Singanamalla}, {Singer}, {Sladen}, {Sooley}, {Sornarajah}, {Streicher}, {Teuben}, {Thomas}, {Tremblay}, {Turner}, {Terr{\'o}n}, {van Kerkwijk}, {de la Vega}, {Watkins}, {Weaver}, {Whitmore}, {Woillez}, {Zabalza}, \& {Astropy Contributors}}]{astropy2}
{Astropy Collaboration}, {Price-Whelan}, A.~M., {Sip{\H{o}}cz}, B.~M., {et~al.} 2018, \bibinfo{title}{{The Astropy Project: Building an Open-science Project and Status of the v2.0 Core Package},} \aj, 156, 123, \dodoi{10.3847/1538-3881/aabc4f}

\bibitem[{ {Astropy Collaboration} {et~al.}(2022){Astropy Collaboration}, {Price-Whelan}, {Lim}, {Earl}, {Starkman}, {Bradley}, {Shupe}, {Patil}, {Corrales}, {Brasseur}, {N{\"o}the}, {Donath}, {Tollerud}, {Morris}, {Ginsburg}, {Vaher}, {Weaver}, {Tocknell}, {Jamieson}, {van Kerkwijk}, {Robitaille}, {Merry}, {Bachetti}, {G{\"u}nther}, {Aldcroft}, {Alvarado-Montes}, {Archibald}, {B{\'o}di}, {Bapat}, {Barentsen}, {Baz{\'a}n}, {Biswas}, {Boquien}, {Burke}, {Cara}, {Cara}, {Conroy}, {Conseil}, {Craig}, {Cross}, {Cruz}, {D'Eugenio}, {Dencheva}, {Devillepoix}, {Dietrich}, {Eigenbrot}, {Erben}, {Ferreira}, {Foreman-Mackey}, {Fox}, {Freij}, {Garg}, {Geda}, {Glattly}, {Gondhalekar}, {Gordon}, {Grant}, {Greenfield}, {Groener}, {Guest}, {Gurovich}, {Handberg}, {Hart}, {Hatfield-Dodds}, {Homeier}, {Hosseinzadeh}, {Jenness}, {Jones}, {Joseph}, {Kalmbach}, {Karamehmetoglu}, {Ka{\l}uszy{\'n}ski}, {Kelley}, {Kern}, {Kerzendorf}, {Koch}, {Kulumani}, {Lee}, {Ly}, {Ma}, {MacBride}, {Maljaars}, {Muna}, {Murphy}, {Norman},
  {O'Steen}, {Oman}, {Pacifici}, {Pascual}, {Pascual-Granado}, {Patil}, {Perren}, {Pickering}, {Rastogi}, {Roulston}, {Ryan}, {Rykoff}, {Sabater}, {Sakurikar}, {Salgado}, {Sanghi}, {Saunders}, {Savchenko}, {Schwardt}, {Seifert-Eckert}, {Shih}, {Jain}, {Shukla}, {Sick}, {Simpson}, {Singanamalla}, {Singer}, {Singhal}, {Sinha}, {Sip{\H{o}}cz}, {Spitler}, {Stansby}, {Streicher}, {{\v{S}}umak}, {Swinbank}, {Taranu}, {Tewary}, {Tremblay}, {de Val-Borro}, {Van Kooten}, {Vasovi{\'c}}, {Verma}, {de Miranda Cardoso}, {Williams}, {Wilson}, {Winkel}, {Wood-Vasey}, {Xue}, {Yoachim}, {Zhang}, {Zonca}, \& {Astropy Project Contributors}}]{astropy3}
{Astropy Collaboration}, {Price-Whelan}, A.~M., {Lim}, P.~L., {et~al.} 2022, \bibinfo{title}{{The Astropy Project: Sustaining and Growing a Community-oriented Open-source Project and the Latest Major Release (v5.0) of the Core Package},} \apj, 935, 167, \dodoi{10.3847/1538-4357/ac7c74}

\bibitem[{H. {Atek} {et~al.}(2024){Atek}, {Labb{\'e}}, {Furtak}, {Chemerynska}, {Fujimoto}, {Setton}, {Miller}, {Oesch}, {Bezanson}, {Price}, {Dayal}, {Zitrin}, {Kokorev}, {Weaver}, {Brammer}, {Dokkum}, {Williams}, {Cutler}, {Feldmann}, {Fudamoto}, {Greene}, {Leja}, {Maseda}, {Muzzin}, {Pan}, {Papovich}, {Nelson}, {Nanayakkara}, {Stark}, {Stefanon}, {Suess}, {Wang}, \& {Whitaker}}]{atek24}
{Atek}, H., {Labb{\'e}}, I., {Furtak}, L.~J., {et~al.} 2024, \bibinfo{title}{{Most of the photons that reionized the Universe came from dwarf galaxies},} \nat, 626, 975, \dodoi{10.1038/s41586-024-07043-6}

\bibitem[{R. {Begley} {et~al.}(2024){Begley}, {Cullen}, {McLure}, {Shapley}, {Dunlop}, {Carnall}, {McLeod}, {Donnan}, {Hamadouche}, \& {Stanton}}]{begley24}
{Begley}, R., {Cullen}, F., {McLure}, R.~J., {et~al.} 2024, \bibinfo{title}{{Connecting the escape fraction of Lyman-alpha and Lyman-continuum photons in star-forming galaxies at z ≃ 4-5},} \mnras, 527, 4040, \dodoi{10.1093/mnras/stad3417}

\bibitem[{S.~E.~I. {Bosman} {et~al.}(2020){Bosman}, {Kakiichi}, {Meyer}, {Gronke}, {Laporte}, \& {Ellis}}]{bosman20}
{Bosman}, S. E.~I., {Kakiichi}, K., {Meyer}, R.~A., {et~al.} 2020, \bibinfo{title}{{Three Ly{\ensuremath{\alpha}} Emitting Galaxies within a Quasar Proximity Zone at z {\ensuremath{\sim}} 5.8},} \apj, 896, 49, \dodoi{10.3847/1538-4357/ab85cd}

\bibitem[{S.~E.~I. {Bosman} {et~al.}(2022){Bosman}, {Davies}, {Becker}, {Keating}, {Davies}, {Zhu}, {Eilers}, {D'Odorico}, {Bian}, {Bischetti}, {Cristiani}, {Fan}, {Farina}, {Haehnelt}, {Hennawi}, {Kulkarni}, {Mesinger}, {Meyer}, {Onoue}, {Pallottini}, {Qin}, {Ryan-Weber}, {Schindler}, {Walter}, {Wang}, \& {Yang}}]{bosman22}
{Bosman}, S. E.~I., {Davies}, F.~B., {Becker}, G.~D., {et~al.} 2022, \bibinfo{title}{{Hydrogen reionization ends by z = 5.3: Lyman-{\ensuremath{\alpha}} optical depth measured by the XQR-30 sample},} \mnras, 514, 55, \dodoi{10.1093/mnras/stac1046}

\bibitem[{R.~J. {Bouwens} {et~al.}(2022){Bouwens}, {Illingworth}, {Ellis}, {Oesch}, \& {Stefanon}}]{bouwens22}
{Bouwens}, R.~J., {Illingworth}, G., {Ellis}, R.~S., {Oesch}, P., \& {Stefanon}, M. 2022, \bibinfo{title}{{z 2-9 Galaxies Magnified by the Hubble Frontier Field Clusters. II. Luminosity Functions and Constraints on a Faint-end Turnover},} \apj, 940, 55, \dodoi{10.3847/1538-4357/ac86d1}

\bibitem[{R.~J. {Bouwens} {et~al.}(2012){Bouwens}, {Illingworth}, {Oesch}, {Trenti}, {Labb{\'e}}, {Franx}, {Stiavelli}, {Carollo}, {van Dokkum}, \& {Magee}}]{bouwens12}
{Bouwens}, R.~J., {Illingworth}, G.~D., {Oesch}, P.~A., {et~al.} 2012, \bibinfo{title}{{Lower-luminosity Galaxies Could Reionize the Universe: Very Steep Faint-end Slopes to the UV Luminosity Functions at z >= 5-8 from the HUDF09 WFC3/IR Observations},} \apjl, 752, L5, \dodoi{10.1088/2041-8205/752/1/L5}

\bibitem[{K.~N.~K. {Boyett} {et~al.}(2022){Boyett}, {Stark}, {Bunker}, {Tang}, \& {Maseda}}]{boyett22}
{Boyett}, K. N.~K., {Stark}, D.~P., {Bunker}, A.~J., {Tang}, M., \& {Maseda}, M.~V. 2022, \bibinfo{title}{{The [O III]{\ensuremath{\lambda}}5007 equivalent width distribution at z 2: the redshift evolution of the extreme emission line galaxies},} \mnras, 513, 4451, \dodoi{10.1093/mnras/stac1109}

\bibitem[{G. {Brammer}(2023){Brammer}}]{msaexp}
{Brammer}, G. 2023, \bibinfo{title}{{msaexp: NIRSpec analyis tools},}, 0.6.17 Zenodo, \dodoi{10.5281/zenodo.8319596}

\bibitem[{Z. {Chen} {et~al.}(2024){Chen}, {Stark}, {Mason}, {Topping}, {Whitler}, {Tang}, {Endsley}, \& {Charlot}}]{chen24}
{Chen}, Z., {Stark}, D.~P., {Mason}, C., {et~al.} 2024, \bibinfo{title}{{JWST spectroscopy of z 5-8 UV-selected galaxies: new constraints on the evolution of the Ly {\ensuremath{\alpha}} escape fraction in the reionization era},} \mnras, 528, 7052, \dodoi{10.1093/mnras/stae455}

\bibitem[{Z. {Chen} {et~al.}(2025){Chen}, {Stark}, {Mason}, {Tang}, {Whitler}, {Lu}, \& {Topping}}]{chen25}
{Chen}, Z., {Stark}, D.~P., {Mason}, C.~A., {et~al.} 2025, \bibinfo{title}{{The Impact of Galaxy Overdensities and Ionized Bubbles on Ly$α$ Emission at $z\sim7.0-8.5$},} arXiv e-prints, arXiv:2505.24080, \dodoi{10.48550/arXiv.2505.24080}

\bibitem[{J. {Chisholm} {et~al.}(2022){Chisholm}, {Saldana-Lopez}, {Flury}, {Schaerer}, {Jaskot}, {Amor{\'\i}n}, {Atek}, {Finkelstein}, {Fleming}, {Ferguson}, {Fern{\'a}ndez}, {Giavalisco}, {Hayes}, {Heckman}, {Henry}, {Ji}, {Marques-Chaves}, {Mauerhofer}, {McCandliss}, {Oey}, {{\"O}stlin}, {Rutkowski}, {Scarlata}, {Thuan}, {Trebitsch}, {Wang}, {Worseck}, \& {Xu}}]{Chisholm22}
{Chisholm}, J., {Saldana-Lopez}, A., {Flury}, S., {et~al.} 2022, \bibinfo{title}{{The far-ultraviolet continuum slope as a Lyman Continuum escape estimator at high redshift},} \mnras, 517, 5104, \dodoi{10.1093/mnras/stac2874}

\bibitem[{F.~B. {Davies}(2020){Davies}}]{davies2020b}
{Davies}, F.~B. 2020, \bibinfo{title}{{Ionization bias and the ghost proximity effect near z {\ensuremath{\gtrsim}} 6 quasars in the shadow of proximate absorption systems},} \mnras, 494, 2937, \dodoi{10.1093/mnras/staa528}

\bibitem[{M. {Dijkstra} {et~al.}(2016){Dijkstra}, {Gronke}, \& {Venkatesan}}]{dijkstra16}
{Dijkstra}, M., {Gronke}, M., \& {Venkatesan}, A. 2016, \bibinfo{title}{{The Ly{\ensuremath{\alpha}}-LyC Connection: Evidence for an Enhanced Contribution of UV-faint Galaxies to Cosmic Reionization},} \apj, 828, 71, \dodoi{10.3847/0004-637X/828/2/71}

\bibitem[{A.-C. {Eilers} {et~al.}(2017){Eilers}, {Davies}, {Hennawi}, {Prochaska}, {Luki{\'c}}, \& {Mazzucchelli}}]{eilers17}
{Eilers}, A.-C., {Davies}, F.~B., {Hennawi}, J.~F., {et~al.} 2017, \bibinfo{title}{{Implications of z {\ensuremath{\sim}} 6 Quasar Proximity Zones for the Epoch of Reionization and Quasar Lifetimes},} \apj, 840, 24, \dodoi{10.3847/1538-4357/aa6c60}

\bibitem[{A.-C. {Eilers} {et~al.}(2025){Eilers}, {Yue}, {Matthee}, {Hennawi}, {Davies}, {Simcoe}, {Teague}, {Bordoloi}, {Brammer}, {Kang}, {Kashino}, {Mackenzie}, {Naidu}, \& {Navarrete}}]{eilers25}
{Eilers}, A.-C., {Yue}, M., {Matthee}, J., {et~al.} 2025, \bibinfo{title}{{The Light Echo of a High-Redshift Quasar mapped with Lyman-$α$ Tomography},} arXiv e-prints, arXiv:2509.05417, \dodoi{10.48550/arXiv.2509.05417}

\bibitem[{R. {Endsley} {et~al.}(2021){Endsley}, {Stark}, {Charlot}, {Chevallard}, {Robertson}, {Bouwens}, \& {Stefanon}}]{endsley21}
{Endsley}, R., {Stark}, D.~P., {Charlot}, S., {et~al.} 2021, \bibinfo{title}{{MMT spectroscopy of Lyman-alpha at z ≃ 7: evidence for accelerated reionization around massive galaxies},} \mnras, 502, 6044, \dodoi{10.1093/mnras/stab432}

\bibitem[{X. {Fan} {et~al.}(2006){Fan}, {Strauss}, {Becker}, {White}, {Gunn}, {Knapp}, {Richards}, {Schneider}, {Brinkmann}, \& {Fukugita}}]{fan06}
{Fan}, X., {Strauss}, M.~A., {Becker}, R.~H., {et~al.} 2006, \bibinfo{title}{{Constraining the Evolution of the Ionizing Background and the Epoch of Reionization with z\raisebox{-0.5ex}\textasciitilde6 Quasars. II. A Sample of 19 Quasars},} \aj, 132, 117, \dodoi{10.1086/504836}

\bibitem[{S.~L. {Finkelstein} {et~al.}(2019){Finkelstein}, {D'Aloisio}, {Paardekooper}, {Ryan}, {Behroozi}, {Finlator}, {Livermore}, {Upton Sanderbeck}, {Dalla Vecchia}, \& {Khochfar}}]{finkelstein19}
{Finkelstein}, S.~L., {D'Aloisio}, A., {Paardekooper}, J.-P., {et~al.} 2019, \bibinfo{title}{{Conditions for Reionizing the Universe with a Low Galaxy Ionizing Photon Escape Fraction},} \apj, 879, 36, \dodoi{10.3847/1538-4357/ab1ea8}

\bibitem[{M.~J. {Hayes} {et~al.}(2021){Hayes}, {Runnholm}, {Gronke}, \& {Scarlata}}]{hayes20}
{Hayes}, M.~J., {Runnholm}, A., {Gronke}, M., \& {Scarlata}, C. 2021, \bibinfo{title}{{Spectral Shapes of the Ly{\ensuremath{\alpha}} Emission from Galaxies. I. Blueshifted Emission and Intrinsic Invariance with Redshift},} \apj, 908, 36, \dodoi{10.3847/1538-4357/abd246}

\bibitem[{Y.~I. {Izotov} {et~al.}(2022){Izotov}, {Chisholm}, {Worseck}, {Guseva}, {Schaerer}, \& {Prochaska}}]{izotov22}
{Izotov}, Y.~I., {Chisholm}, J., {Worseck}, G., {et~al.} 2022, \bibinfo{title}{{Lyman alpha and Lyman continuum emission of Mg II-selected star-forming galaxies},} \mnras, 515, 2864, \dodoi{10.1093/mnras/stac1899}

\bibitem[{Y.~I. {Izotov} {et~al.}(2018){Izotov}, {Worseck}, {Schaerer}, {Guseva}, {Thuan}, {Fricke}, \& {Orlitov{\'a}}}]{izotov18}
{Izotov}, Y.~I., {Worseck}, G., {Schaerer}, D., {et~al.} 2018, \bibinfo{title}{{Low-redshift Lyman continuum leaking galaxies with high [O III]/[O II] ratios},} \mnras, 478, 4851, \dodoi{10.1093/mnras/sty1378}

\bibitem[{A.~E. {Jaskot} {et~al.}(2024){Jaskot}, {Silveyra}, {Plantinga}, {Flury}, {Hayes}, {Chisholm}, {Heckman}, {Pentericci}, {Schaerer}, {Trebitsch}, {Verhamme}, {Carr}, {Ferguson}, {Ji}, {Giavalisco}, {Henry}, {Marques-Chaves}, {{\"O}stlin}, {Saldana-Lopez}, {Scarlata}, {Worseck}, \& {Xu}}]{jaskot2024}
{Jaskot}, A.~E., {Silveyra}, A.~C., {Plantinga}, A., {et~al.} 2024, \bibinfo{title}{{Multivariate Predictors of Lyman Continuum Escape. II. Predicting Lyman Continuum Escape Fractions for High-redshift Galaxies},} \apj, 973, 111, \dodoi{10.3847/1538-4357/ad5557}

\bibitem[{X. {Jin} {et~al.}(2023){Jin}, {Yang}, {Fan}, {Wang}, {Ba{\~n}ados}, {Bian}, {Davies}, {Eilers}, {Farina}, {Hennawi}, {Pacucci}, {Venemans}, \& {Walter}}]{jin23}
{Jin}, X., {Yang}, J., {Fan}, X., {et~al.} 2023, \bibinfo{title}{{(Nearly) Model-independent Constraints on the Neutral Hydrogen Fraction in the Intergalactic Medium at z 5-7 Using Dark Pixel Fractions in Ly{\ensuremath{\alpha}} and Ly{\ensuremath{\beta}} Forests},} \apj, 942, 59, \dodoi{10.3847/1538-4357/aca678}

\bibitem[{Y. {Kageura} {et~al.}(2025){Kageura}, {Ouchi}, {Nakane}, {Umeda}, {Harikane}, {Yoshiura}, {Nakajima}, {Yajima}, \& {Thai}}]{kageura25}
{Kageura}, Y., {Ouchi}, M., {Nakane}, M., {et~al.} 2025, \bibinfo{title}{{Census of Ly$\alpha$ Emission from $\sim 600$ Galaxies at $z=5-14$: Evolution of the Ly$\alpha$ Luminosity Function and a Late Sharp Cosmic Reionization},} arXiv e-prints, arXiv:2501.05834, \dodoi{10.48550/arXiv.2501.05834}

\bibitem[{D. {Kashino} {et~al.}(2023){Kashino}, {Lilly}, {Matthee}, {Eilers}, {Mackenzie}, {Bordoloi}, \& {Simcoe}}]{kashino23}
{Kashino}, D., {Lilly}, S.~J., {Matthee}, J., {et~al.} 2023, \bibinfo{title}{{EIGER. I. A Large Sample of [O III]-emitting Galaxies at 5.3 < z < 6.9 and Direct Evidence for Local Reionization by Galaxies},} \apj, 950, 66, \dodoi{10.3847/1538-4357/acc588}

\bibitem[{D. {Kashino} {et~al.}(2025){Kashino}, {Lilly}, {Matthee}, {Mackenzie}, {Eilers}, {Bordoloi}, {Simcoe}, {Naidu}, {Yue}, \& {Liu}}]{kashino25}
{Kashino}, D., {Lilly}, S.~J., {Matthee}, J., {et~al.} 2025, \bibinfo{title}{{EIGER VII. The evolving relationship between galaxies and the intergalactic medium in the final stages of reionization},} arXiv e-prints, arXiv:2506.03121, \dodoi{10.48550/arXiv.2506.03121}

\bibitem[{I.~S. {Khrykin} {et~al.}(2021){Khrykin}, {Hennawi}, {Worseck}, \& {Davies}}]{Khrykin21}
{Khrykin}, I.~S., {Hennawi}, J.~F., {Worseck}, G., \& {Davies}, F.~B. 2021, \bibinfo{title}{{The first measurement of the quasar lifetime distribution},} \mnras, 505, 649, \dodoi{10.1093/mnras/stab1288}

\bibitem[{X. {Lin} {et~al.}(2024){Lin}, {Cai}, {Wu}, {Li}, {Sun}, {Fan}, {Chen}, {Li}, {Bian}, {Ning}, {Jiang}, {Bruzual}, {Charlot}, \& {Chevallard}}]{lin24}
{Lin}, X., {Cai}, Z., {Wu}, Y., {et~al.} 2024, \bibinfo{title}{{Quantifying the Escape of Ly{\ensuremath{\alpha}} at z {\ensuremath{\approx}} 5{\textendash}6: A Census of Ly{\ensuremath{\alpha}} Escape Fraction with H{\ensuremath{\alpha}}-emitting Galaxies Spectroscopically Confirmed by JWST and VLT/MUSE},} \apjs, 272, 33, \dodoi{10.3847/1538-4365/ad3e7d}

\bibitem[{T.-Y. {Lu} {et~al.}(2024){Lu}, {Mason}, {Mesinger}, {Prelogovi{\'c}}, {Nikoli{\'c}}, {Hutter}, {Gagnon-Hartman}, {Tang}, {Qin}, \& {Kakiichi}}]{lu24}
{Lu}, T.-Y., {Mason}, C.~A., {Mesinger}, A., {et~al.} 2024, \bibinfo{title}{{Mapping reionization bubbles in the JWST era I: empirical edge detection with Lyman alpha emission from galaxies},} arXiv e-prints, arXiv:2411.04176, \dodoi{10.48550/arXiv.2411.04176}

\bibitem[{X. {Ma} {et~al.}(2016){Ma}, {Hopkins}, {Kasen}, {Quataert}, {Faucher-Gigu{\`e}re}, {Kere{\v{s}}}, {Murray}, \& {Strom}}]{ma16}
{Ma}, X., {Hopkins}, P.~F., {Kasen}, D., {et~al.} 2016, \bibinfo{title}{{Binary stars can provide the `missing photons' needed for reionization},} \mnras, 459, 3614, \dodoi{10.1093/mnras/stw941}

\bibitem[{X. {Ma} {et~al.}(2015){Ma}, {Kasen}, {Hopkins}, {Faucher-Gigu{\`e}re}, {Quataert}, {Kere{\v{s}}}, \& {Murray}}]{ma15}
{Ma}, X., {Kasen}, D., {Hopkins}, P.~F., {et~al.} 2015, \bibinfo{title}{{The difficulty of getting high escape fractions of ionizing photons from high-redshift galaxies: a view from the FIRE cosmological simulations},} \mnras, 453, 960, \dodoi{10.1093/mnras/stv1679}

\bibitem[{S. {Mascia} {et~al.}(2023){Mascia}, {Pentericci}, {Calabr{\`o}}, {Treu}, {Santini}, {Yang}, {Napolitano}, {Roberts-Borsani}, {Bergamini}, {Grillo}, {Rosati}, {Vulcani}, {Castellano}, {Boyett}, {Fontana}, {Glazebrook}, {Henry}, {Mason}, {Merlin}, {Morishita}, {Nanayakkara}, {Paris}, {Roy}, {Williams}, {Wang}, {Brammer}, {Brada{\v{c}}}, {Chen}, {Kelly}, {Koekemoer}, {Trenti}, \& {Windhorst}}]{mascia23}
{Mascia}, S., {Pentericci}, L., {Calabr{\`o}}, A., {et~al.} 2023, \bibinfo{title}{{Closing in on the sources of cosmic reionization: First results from the GLASS-JWST program},} \aap, 672, A155, \dodoi{10.1051/0004-6361/202345866}

\bibitem[{C.~A. {Mason} {et~al.}(2018){Mason}, {Treu}, {Dijkstra}, {Mesinger}, {Trenti}, {Pentericci}, {de Barros}, \& {Vanzella}}]{mason18}
{Mason}, C.~A., {Treu}, T., {Dijkstra}, M., {et~al.} 2018, \bibinfo{title}{{The Universe Is Reionizing at z {\ensuremath{\sim}} 7: Bayesian Inference of the IGM Neutral Fraction Using Ly{\ensuremath{\alpha}} Emission from Galaxies},} \apj, 856, 2, \dodoi{10.3847/1538-4357/aab0a7}

\bibitem[{C.~A. {Mason} {et~al.}(2019){Mason}, {Fontana}, {Treu}, {Schmidt}, {Hoag}, {Abramson}, {Amorin}, {Brada{\v{c}}}, {Guaita}, {Jones}, {Henry}, {Malkan}, {Pentericci}, {Trenti}, \& {Vanzella}}]{mason19}
{Mason}, C.~A., {Fontana}, A., {Treu}, T., {et~al.} 2019, \bibinfo{title}{{Inferences on the timeline of reionization at z {\ensuremath{\sim}} 8 from the KMOS Lens-Amplified Spectroscopic Survey},} \mnras, 485, 3947, \dodoi{10.1093/mnras/stz632}

\bibitem[{J. {Matthee} {et~al.}(2023){Matthee}, {Mackenzie}, {Simcoe}, {Kashino}, {Lilly}, {Bordoloi}, \& {Eilers}}]{matthee23}
{Matthee}, J., {Mackenzie}, R., {Simcoe}, R.~A., {et~al.} 2023, \bibinfo{title}{{EIGER. II. First Spectroscopic Characterization of the Young Stars and Ionized Gas Associated with Strong H{\ensuremath{\beta}} and [O III] Line Emission in Galaxies at z = 5-7 with JWST},} \apj, 950, 67, \dodoi{10.3847/1538-4357/acc846}

\bibitem[{J. {Matthee} {et~al.}(2018){Matthee}, {Sobral}, {Gronke}, {Paulino-Afonso}, {Stefanon}, \& {R{\"o}ttgering}}]{matthee18}
{Matthee}, J., {Sobral}, D., {Gronke}, M., {et~al.} 2018, \bibinfo{title}{{Confirmation of double peaked Ly{\ensuremath{\alpha}} emission at z = 6.593. Witnessing a galaxy directly contributing to the reionisation of the Universe},} \aap, 619, A136, \dodoi{10.1051/0004-6361/201833528}

\bibitem[{J. {Matthee} {et~al.}(2022){Matthee}, {Naidu}, {Pezzulli}, {Gronke}, {Sobral}, {Oesch}, {Hayes}, {Erb}, {Schaerer}, {Amor{\'\i}n}, {Tacchella}, {Paulino-Afonso}, {Llerena}, {Calhau}, \& {R{\"o}ttgering}}]{matthee22}
{Matthee}, J., {Naidu}, R.~P., {Pezzulli}, G., {et~al.} 2022, \bibinfo{title}{{(Re)Solving reionization with Ly{\ensuremath{\alpha}}: how bright Ly{\ensuremath{\alpha}} Emitters account for the z {\ensuremath{\approx}} 2-8 cosmic ionizing background},} \mnras, 512, 5960, \dodoi{10.1093/mnras/stac801}

\bibitem[{C. {Mazzucchelli} {et~al.}(2023){Mazzucchelli}, {Bischetti}, {D'Odorico}, {Feruglio}, {Schindler}, {Onoue}, {Ba{\~n}ados}, {Becker}, {Bian}, {Carniani}, {Decarli}, {Eilers}, {Farina}, {Gallerani}, {Lai}, {Meyer}, {Rojas-Ruiz}, {Satyavolu}, {Venemans}, {Wang}, {Yang}, \& {Zhu}}]{Mazzucchelli23}
{Mazzucchelli}, C., {Bischetti}, M., {D'Odorico}, V., {et~al.} 2023, \bibinfo{title}{{XQR-30: Black hole masses and accretion rates of 42 z {\ensuremath{\gtrsim}} 6 quasars},} \aap, 676, A71, \dodoi{10.1051/0004-6361/202346317}

\bibitem[{R.~A. {Meyer} {et~al.}(2025){Meyer}, {Roberts-Borsani}, {Oesch}, \& {Ellis}}]{meyer25}
{Meyer}, R.~A., {Roberts-Borsani}, G., {Oesch}, P., \& {Ellis}, R.~S. 2025, \bibinfo{title}{{Probing patchy reionisation with JWST: IGM opacity constraints from the Lyman-$α$ forest of galaxies in legacy extragalactic fields},} arXiv e-prints, arXiv:2504.02683, \dodoi{10.48550/arXiv.2504.02683}

\bibitem[{R.~P. {Naidu} {et~al.}(2018){Naidu}, {Forrest}, {Oesch}, {Tran}, \& {Holden}}]{naidu18}
{Naidu}, R.~P., {Forrest}, B., {Oesch}, P.~A., {Tran}, K.-V.~H., \& {Holden}, B.~P. 2018, \bibinfo{title}{{A low Lyman Continuum escape fraction of <10 per cent for extreme [O III] emitters in an overdensity at z {\ensuremath{\sim}} 3.5},} \mnras, 478, 791, \dodoi{10.1093/mnras/sty961}

\bibitem[{R.~P. {Naidu} {et~al.}(2020){Naidu}, {Tacchella}, {Mason}, {Bose}, {Oesch}, \& {Conroy}}]{naidu20}
{Naidu}, R.~P., {Tacchella}, S., {Mason}, C.~A., {et~al.} 2020, \bibinfo{title}{{Rapid Reionization by the Oligarchs: The Case for Massive, UV-bright, Star-forming Galaxies with High Escape Fractions},} \apj, 892, 109, \dodoi{10.3847/1538-4357/ab7cc9}

\bibitem[{R.~P. {Naidu} {et~al.}(2022){Naidu}, {Matthee}, {Oesch}, {Conroy}, {Sobral}, {Pezzulli}, {Hayes}, {Erb}, {Amor{\'\i}n}, {Gronke}, {Schaerer}, {Tacchella}, {Kerutt}, {Paulino-Afonso}, {Calhau}, {Llerena}, \& {R{\"o}ttgering}}]{naidu22}
{Naidu}, R.~P., {Matthee}, J., {Oesch}, P.~A., {et~al.} 2022, \bibinfo{title}{{The synchrony of production and escape: half the bright Ly{\ensuremath{\alpha}} emitters at z {\ensuremath{\approx}} 2 have Lyman continuum escape fractions {\ensuremath{\approx}}50 per cent},} \mnras, 510, 4582, \dodoi{10.1093/mnras/stab3601}

\bibitem[{K. {Nakajima} {et~al.}(2018){Nakajima}, {Schaerer}, {Le F{\`e}vre}, {Amor{\'\i}n}, {Talia}, {Lemaux}, {Tasca}, {Vanzella}, {Zamorani}, {Bardelli}, {Grazian}, {Guaita}, {Hathi}, {Pentericci}, \& {Zucca}}]{nakajima18}
{Nakajima}, K., {Schaerer}, D., {Le F{\`e}vre}, O., {et~al.} 2018, \bibinfo{title}{{The VIMOS Ultra Deep Survey: Nature, ISM properties, and ionizing spectra of CIII]{\ensuremath{\lambda}}1909 emitters at z = 2-4},} \aap, 612, A94, \dodoi{10.1051/0004-6361/201731935}

\bibitem[{I. {Nikoli{\'c}} {et~al.}(2025){Nikoli{\'c}}, {Mesinger}, {Mason}, {Lu}, {Tang}, {Prelogovi{\'c}}, {Gagnon-Hartman}, \& {Stark}}]{nikolic25}
{Nikoli{\'c}}, I., {Mesinger}, A., {Mason}, C.~A., {et~al.} 2025, \bibinfo{title}{{Mapping reionization bubbles in the JWST era II: inferring the position and characteristic size of individual bubbles},} arXiv e-prints, arXiv:2501.07980, \dodoi{10.48550/arXiv.2501.07980}

\bibitem[{J.~B. {Oke} \& J.~E. {Gunn}(1983){Oke} \& {Gunn}}]{ABsystem}
{Oke}, J.~B., \& {Gunn}, J.~E. 1983, \bibinfo{title}{{Secondary standard stars for absolute spectrophotometry.},} \apj, 266, 713, \dodoi{10.1086/160817}

\bibitem[{C. {Papovich} {et~al.}(2025){Papovich}, {Cole}, {Hu}, {Finkelstein}, {Shen}, {Arrabal Haro}, {Amor{\'\i}n}, {Backhaus}, {Bagley}, {Bhatawdekar}, {Calabr{\'o}}, {Carnall}, {Cleri}, {Daddi}, {Dickinson}, {Grogin}, {Holwerda}, {Jaskot}, {Koekemoer}, {Llerena}, {Lucas}, {Mascia}, {Pacucci}, {Pentericci}, {P{\'e}rez-Gonz{\'a}lez}, {Pirzkal}, {Raghunathan}, {Seill{\'e}}, {Somerville}, \& {Yung}}]{papovich25}
{Papovich}, C., {Cole}, J.~W., {Hu}, W., {et~al.} 2025, \bibinfo{title}{{Galaxies in the Epoch of Reionization Are All Bark and No Bite -- Plenty of Ionizing Photons, Low Escape Fractions},} arXiv e-prints, arXiv:2505.08870, \dodoi{10.48550/arXiv.2505.08870}

\bibitem[{ {Planck Collaboration} {et~al.}(2020){Planck Collaboration}, {Aghanim}, {Akrami}, {Ashdown}, {Aumont}, {Baccigalupi}, {Ballardini}, {Banday}, {Barreiro}, {Bartolo}, {Basak}, {Battye}, {Benabed}, {Bernard}, {Bersanelli}, {Bielewicz}, {Bock}, {Bond}, {Borrill}, {Bouchet}, {Boulanger}, {Bucher}, {Burigana}, {Butler}, {Calabrese}, {Cardoso}, {Carron}, {Challinor}, {Chiang}, {Chluba}, {Colombo}, {Combet}, {Contreras}, {Crill}, {Cuttaia}, {de Bernardis}, {de Zotti}, {Delabrouille}, {Delouis}, {Di Valentino}, {Diego}, {Dor{\'e}}, {Douspis}, {Ducout}, {Dupac}, {Dusini}, {Efstathiou}, {Elsner}, {En{\ss}lin}, {Eriksen}, {Fantaye}, {Farhang}, {Fergusson}, {Fernandez-Cobos}, {Finelli}, {Forastieri}, {Frailis}, {Fraisse}, {Franceschi}, {Frolov}, {Galeotta}, {Galli}, {Ganga}, {G{\'e}nova-Santos}, {Gerbino}, {Ghosh}, {Gonz{\'a}lez-Nuevo}, {G{\'o}rski}, {Gratton}, {Gruppuso}, {Gudmundsson}, {Hamann}, {Handley}, {Hansen}, {Herranz}, {Hildebrandt}, {Hivon}, {Huang}, {Jaffe}, {Jones}, {Karakci}, {Keih{\"a}nen},
  {Keskitalo}, {Kiiveri}, {Kim}, {Kisner}, {Knox}, {Krachmalnicoff}, {Kunz}, {Kurki-Suonio}, {Lagache}, {Lamarre}, {Lasenby}, {Lattanzi}, {Lawrence}, {Le Jeune}, {Lemos}, {Lesgourgues}, {Levrier}, {Lewis}, {Liguori}, {Lilje}, {Lilley}, {Lindholm}, {L{\'o}pez-Caniego}, {Lubin}, {Ma}, {Mac{\'\i}as-P{\'e}rez}, {Maggio}, {Maino}, {Mandolesi}, {Mangilli}, {Marcos-Caballero}, {Maris}, {Martin}, {Martinelli}, {Mart{\'\i}nez-Gonz{\'a}lez}, {Matarrese}, {Mauri}, {McEwen}, {Meinhold}, {Melchiorri}, {Mennella}, {Migliaccio}, {Millea}, {Mitra}, {Miville-Desch{\^e}nes}, {Molinari}, {Montier}, {Morgante}, {Moss}, {Natoli}, {N{\o}rgaard-Nielsen}, {Pagano}, {Paoletti}, {Partridge}, {Patanchon}, {Peiris}, {Perrotta}, {Pettorino}, {Piacentini}, {Polastri}, {Polenta}, {Puget}, {Rachen}, {Reinecke}, {Remazeilles}, {Renzi}, {Rocha}, {Rosset}, {Roudier}, {Rubi{\~n}o-Mart{\'\i}n}, {Ruiz-Granados}, {Salvati}, {Sandri}, {Savelainen}, {Scott}, {Shellard}, {Sirignano}, {Sirri}, {Spencer}, {Sunyaev}, {Suur-Uski}, {Tauber}, {Tavagnacco},
  {Tenti}, {Toffolatti}, {Tomasi}, {Trombetti}, {Valenziano}, {Valiviita}, {Van Tent}, {Vibert}, {Vielva}, {Villa}, {Vittorio}, {Wandelt}, {Wehus}, {White}, {White}, {Zacchei}, \& {Zonca}}]{cmb}
{Planck Collaboration}, {Aghanim}, N., {Akrami}, Y., {et~al.} 2020, \bibinfo{title}{{Planck 2018 results. VI. Cosmological parameters},} \aap, 641, A6, \dodoi{10.1051/0004-6361/201833910}

\bibitem[{J.~X. Prochaska {et~al.}(2020)Prochaska, Hennawi, Westfall, Cooke, Wang, Hsyu, Davies, Farina, \& Pelliccia}]{pypeit}
Prochaska, J.~X., Hennawi, J.~F., Westfall, K.~B., {et~al.} 2020, \bibinfo{title}{PypeIt: The Python Spectroscopic Data Reduction Pipeline,} Journal of Open Source Software, 5, 2308, \dodoi{10.21105/joss.02308}

\bibitem[{K. {Protu{\v{s}}ov{\'a}} {et~al.}(2024){Protu{\v{s}}ov{\'a}}, {Bosman}, {Wang}, {Meyer}, {Champagne}, {Davies}, {Eilers}, {Fan}, {Hennawi}, {Jin}, {Jun}, {Kakiichi}, {Li}, {Liu}, \& {Yang}}]{protusova24}
{Protu{\v{s}}ov{\'a}}, K., {Bosman}, S. E.~I., {Wang}, F., {et~al.} 2024, \bibinfo{title}{{A unique window into the Epoch of Reionisation: A double-peaked Lyman-$\alpha$ emitter in the proximity zone of a quasar at $z\sim 6.6$},} arXiv e-prints, arXiv:2412.12256, \dodoi{10.48550/arXiv.2412.12256}

\bibitem[{B.~E. {Robertson}(2022){Robertson}}]{robertson22}
{Robertson}, B.~E. 2022, \bibinfo{title}{{Galaxy Formation and Reionization: Key Unknowns and Expected Breakthroughs by the James Webb Space Telescope},} \araa, 60, 121, \dodoi{10.1146/annurev-astro-120221-044656}

\bibitem[{J. {Rosdahl} {et~al.}(2022){Rosdahl}, {Blaizot}, {Katz}, {Kimm}, {Garel}, {Haehnelt}, {Keating}, {Martin-Alvarez}, {Michel-Dansac}, \& {Ocvirk}}]{Rosdahl22}
{Rosdahl}, J., {Blaizot}, J., {Katz}, H., {et~al.} 2022, \bibinfo{title}{{LyC escape from SPHINX galaxies in the Epoch of Reionization},} \mnras, 515, 2386, \dodoi{10.1093/mnras/stac1942}

\bibitem[{A. {Saxena} {et~al.}(2022){Saxena}, {Cryer}, {Ellis}, {Pentericci}, {Calabr{\`o}}, {Mascia}, {Saldana-Lopez}, {Schaerer}, {Katz}, {Llerena}, \& {Amor{\'\i}n}}]{saxena22}
{Saxena}, A., {Cryer}, E., {Ellis}, R.~S., {et~al.} 2022, \bibinfo{title}{{Strong C IV emission from star-forming galaxies: a case for high Lyman continuum photon escape},} \mnras, 517, 1098, \dodoi{10.1093/mnras/stac2742}

\bibitem[{D. {Schaerer} {et~al.}(2022){Schaerer}, {Izotov}, {Worseck}, {Berg}, {Chisholm}, {Jaskot}, {Nakajima}, {Ravindranath}, {Thuan}, \& {Verhamme}}]{Schaerer22}
{Schaerer}, D., {Izotov}, Y.~I., {Worseck}, G., {et~al.} 2022, \bibinfo{title}{{Strong Lyman continuum emitting galaxies show intense C IV {\ensuremath{\lambda}}1550 emission},} \aap, 658, L11, \dodoi{10.1051/0004-6361/202243149}

\bibitem[{M.~A. {Schenker} {et~al.}(2014){Schenker}, {Ellis}, {Konidaris}, \& {Stark}}]{schenker14}
{Schenker}, M.~A., {Ellis}, R.~S., {Konidaris}, N.~P., \& {Stark}, D.~P. 2014, \bibinfo{title}{{Line-emitting Galaxies beyond a Redshift of 7: An Improved Method for Estimating the Evolving Neutrality of the Intergalactic Medium},} \apj, 795, 20, \dodoi{10.1088/0004-637X/795/1/20}

\bibitem[{M. {Sharma} {et~al.}(2016){Sharma}, {Theuns}, {Frenk}, {Bower}, {Crain}, {Schaller}, \& {Schaye}}]{sharma16}
{Sharma}, M., {Theuns}, T., {Frenk}, C., {et~al.} 2016, \bibinfo{title}{{The brighter galaxies reionized the Universe},} \mnras, 458, L94, \dodoi{10.1093/mnrasl/slw021}

\bibitem[{C. {Simmonds} {et~al.}(2024){Simmonds}, {Tacchella}, {Hainline}, {Johnson}, {Pusk{\'a}s}, {Robertson}, {Baker}, {Bhatawdekar}, {Boyett}, {Bunker}, {Cargile}, {Carniani}, {Chevallard}, {Curti}, {Curtis-Lake}, {Ji}, {Jones}, {Kumari}, {Laseter}, {Maiolino}, {Maseda}, {Rinaldi}, {Stoffers}, {{\"U}bler}, {Villanueva}, {Williams}, {Willott}, {Witstok}, \& {Zhu}}]{simmons24}
{Simmonds}, C., {Tacchella}, S., {Hainline}, K., {et~al.} 2024, \bibinfo{title}{{Ionizing properties of galaxies in JADES for a stellar mass complete sample: resolving the cosmic ionizing photon budget crisis at the Epoch of Reionization},} \mnras, 535, 2998, \dodoi{10.1093/mnras/stae2537}

\bibitem[{D. {Sobral} \& J. {Matthee}(2019){Sobral} \& {Matthee}}]{sobral19}
{Sobral}, D., \& {Matthee}, J. 2019, \bibinfo{title}{{Predicting Ly{\ensuremath{\alpha}} escape fractions with a simple observable. Ly{\ensuremath{\alpha}} in emission as an empirically calibrated star formation rate indicator},} \aap, 623, A157, \dodoi{10.1051/0004-6361/201833075}

\bibitem[{C.~C. {Steidel} {et~al.}(2018){Steidel}, {Bogosavljevi{\'c}}, {Shapley}, {Reddy}, {Rudie}, {Pettini}, {Trainor}, \& {Strom}}]{steidel18}
{Steidel}, C.~C., {Bogosavljevi{\'c}}, M., {Shapley}, A.~E., {et~al.} 2018, \bibinfo{title}{{The Keck Lyman Continuum Spectroscopic Survey (KLCS): The Emergent Ionizing Spectrum of Galaxies at z {\ensuremath{\sim}} 3},} \apj, 869, 123, \dodoi{10.3847/1538-4357/aaed28}

\bibitem[{M. {Tang} {et~al.}(2024{\natexlab{a}}){Tang}, {Stark}, {Topping}, {Mason}, \& {Ellis}}]{tang24b}
{Tang}, M., {Stark}, D.~P., {Topping}, M.~W., {Mason}, C., \& {Ellis}, R.~S. 2024{\natexlab{a}}, \bibinfo{title}{{JWST/NIRSpec Observations of Lyman {\ensuremath{\alpha}} Emission in Star-forming Galaxies at 6.5 {\ensuremath{\lesssim}} z {\ensuremath{\lesssim}} 13},} \apj, 975, 208, \dodoi{10.3847/1538-4357/ad7eb7}

\bibitem[{M. {Tang} {et~al.}(2024{\natexlab{b}}){Tang}, {Stark}, {Ellis}, {Sun}, {Topping}, {Robertson}, {Tacchella}, {Arribas}, {Baker}, {Bhatawdekar}, {Boyett}, {Bunker}, {Charlot}, {Chen}, {Chevallard}, {Jones}, {Kumari}, {Lyu}, {Maiolino}, {Maseda}, {Saxena}, {Whitler}, {Williams}, {Willott}, \& {Witstok}}]{tang24a}
{Tang}, M., {Stark}, D.~P., {Ellis}, R.~S., {et~al.} 2024{\natexlab{b}}, \bibinfo{title}{{Ly{\ensuremath{\alpha}} emission in galaxies at z ≃ 5-6: new insight from JWST into the statistical distributions of Ly{\ensuremath{\alpha}} properties at the end of reionization},} \mnras, 531, 2701, \dodoi{10.1093/mnras/stae1338}

\bibitem[{R. {Thomas} {et~al.}(2021){Thomas}, {Pentericci}, {Le F{\`e}vre}, {Koekemoer}, {Castellano}, {Cimatti}, {Fontanot}, {Gargiulo}, {Garilli}, {Talia}, {Amor{\'\i}n}, {Bardelli}, {Cristiani}, {Cresci}, {Franco}, {Fynbo}, {Hathi}, {Hibon}, {Khusanova}, {Le Brun}, {Lemaux}, {Mannucci}, {Schaerer}, {Zamorani}, \& {Zucca}}]{thomas21}
{Thomas}, R., {Pentericci}, L., {Le F{\`e}vre}, O., {et~al.} 2021, \bibinfo{title}{{Less and more IGM-transmitted galaxies from z {\ensuremath{\sim}} 2.7 to z {\ensuremath{\sim}} 6 from VANDELS and VUDS},} \aap, 650, A63, \dodoi{10.1051/0004-6361/202038438}

\bibitem[{A. {Torralba} {et~al.}(2024){Torralba}, {Matthee}, {Naidu}, {Mackenzie}, {Pezzulli}, {Hutter}, {Arnalte-Mur}, {Gurung-L{\'o}pez}, {Tacchella}, {Oesch}, {Kashino}, {Conroy}, \& {Sobral}}]{Torralba24}
{Torralba}, A., {Matthee}, J., {Naidu}, R.~P., {et~al.} 2024, \bibinfo{title}{{Anatomy of an ionized bubble: NIRCam grism spectroscopy of the z = 6.6 double-peaked Lyman-{\ensuremath{\alpha}} emitter COLA1 and its environment},} \aap, 689, A44, \dodoi{10.1051/0004-6361/202450318}

\bibitem[{H. {Umeda} {et~al.}(2025){Umeda}, {Ouchi}, {Kageura}, {Harikane}, {Nakane}, {Thai}, \& {Nakajima}}]{umeda25}
{Umeda}, H., {Ouchi}, M., {Kageura}, Y., {et~al.} 2025, \bibinfo{title}{{Probing the Cosmic Reionization History with JWST: Gunn-Peterson and Ly$α$ Damping Wing Absorption at $4.5 < z < 13$},} arXiv e-prints, arXiv:2504.04683, \dodoi{10.48550/arXiv.2504.04683}

\bibitem[{D. {{\v{D}}urov{\v{c}}{\'\i}kov{\'a}} {et~al.}(2024){{\v{D}}urov{\v{c}}{\'\i}kov{\'a}}, {Eilers}, {Chen}, {Satyavolu}, {Kulkarni}, {Simcoe}, {Keating}, {Haehnelt}, \& {Ba{\~n}ados}}]{durovcikova24}
{{\v{D}}urov{\v{c}}{\'\i}kov{\'a}}, D., {Eilers}, A.-C., {Chen}, H., {et~al.} 2024, \bibinfo{title}{{Chronicling the Reionization History at 6 {\ensuremath{\lesssim}} z {\ensuremath{\lesssim}} 7 with Emergent Quasar Damping Wings},} \apj, 969, 162, \dodoi{10.3847/1538-4357/ad4888}

\bibitem[{A. {Verhamme} {et~al.}(2017){Verhamme}, {Orlitov{\'a}}, {Schaerer}, {Izotov}, {Worseck}, {Thuan}, \& {Guseva}}]{verhamme16}
{Verhamme}, A., {Orlitov{\'a}}, I., {Schaerer}, D., {et~al.} 2017, \bibinfo{title}{{Lyman-{\ensuremath{\alpha}} spectral properties of five newly discovered Lyman continuum emitters},} \aap, 597, A13, \dodoi{10.1051/0004-6361/201629264}

\bibitem[{A. {Verhamme} {et~al.}(2018){Verhamme}, {Garel}, {Ventou}, {Contini}, {Bouch{\'e}}, {Herenz}, {Richard}, {Bacon}, {Schmidt}, {Maseda}, {Marino}, {Brinchmann}, {Cantalupo}, {Caruana}, {Cl{\'e}ment}, {Diener}, {Drake}, {Hashimoto}, {Inami}, {Kerutt}, {Kollatschny}, {Leclercq}, {Patr{\'\i}cio}, {Schaye}, {Wisotzki}, \& {Zabl}}]{verhamme18}
{Verhamme}, A., {Garel}, T., {Ventou}, E., {et~al.} 2018, \bibinfo{title}{{Recovering the systemic redshift of galaxies from their Lyman alpha line profile},} \mnras, 478, L60, \dodoi{10.1093/mnrasl/sly058}

\bibitem[{L. {Whitler} {et~al.}(2024){Whitler}, {Stark}, {Endsley}, {Chen}, {Mason}, {Topping}, \& {Charlot}}]{whitler24}
{Whitler}, L., {Stark}, D.~P., {Endsley}, R., {et~al.} 2024, \bibinfo{title}{{Insight from JWST/Near Infrared Camera into galaxy overdensities around bright Lyman-alpha emitters during reionization: implications for ionized bubbles at z 9},} \mnras, 529, 855, \dodoi{10.1093/mnras/stae516}

\bibitem[{I.~G.~B. {Wold} {et~al.}(2022){Wold}, {Malhotra}, {Rhoads}, {Wang}, {Hu}, {Perez}, {Zheng}, {Khostovan}, {Walker}, {Barrientos}, {Gonz{\'a}lez-L{\'o}pez}, {Harish}, {Infante}, {Jiang}, {Pharo}, {Moya-Sierralta}, {Bauer}, {Galaz}, {Valdes}, \& {Yang}}]{wold22}
{Wold}, I. G.~B., {Malhotra}, S., {Rhoads}, J., {et~al.} 2022, \bibinfo{title}{{LAGER Ly{\ensuremath{\alpha}} Luminosity Function at z 7: Implications for Reionization},} \apj, 927, 36, \dodoi{10.3847/1538-4357/ac4997}

\bibitem[{J. {Yang} {et~al.}(2020){Yang}, {Wang}, {Fan}, {Hennawi}, {Davies}, {Yue}, {Eilers}, {Farina}, {Wu}, {Bian}, {Pacucci}, \& {Lee}}]{yang20b}
{Yang}, J., {Wang}, F., {Fan}, X., {et~al.} 2020, \bibinfo{title}{{Measurements of the z {\ensuremath{\sim}} 6 Intergalactic Medium Optical Depth and Transmission Spikes Using a New z > 6.3 Quasar Sample},} \apj, 904, 26, \dodoi{10.3847/1538-4357/abbc1b}

\end{thebibliography}
\bibliographystyle{aasjournalv7}
\end{document}